\PassOptionsToPackage{unicode}{hyperref}
\PassOptionsToPackage{hyphens}{url}
\documentclass[
]{statsoc}
\author{}
\date{\vspace{-2.5em}}

\usepackage{amsmath,amssymb}
\usepackage{lmodern}
\usepackage{iftex}
\ifPDFTeX
  \usepackage[T1]{fontenc}
  \usepackage[utf8]{inputenc}
  \usepackage{textcomp} 
\else 
  \usepackage{unicode-math}
  \defaultfontfeatures{Scale=MatchLowercase}
  \defaultfontfeatures[\rmfamily]{Ligatures=TeX,Scale=1}
\fi
\IfFileExists{upquote.sty}{\usepackage{upquote}}{}
\IfFileExists{microtype.sty}{
  \usepackage[]{microtype}
  \UseMicrotypeSet[protrusion]{basicmath} 
}{}
\makeatletter
\@ifundefined{KOMAClassName}{
  \IfFileExists{parskip.sty}{%
    \usepackage{parskip}
  }{
    \setlength{\parindent}{0pt}
    \setlength{\parskip}{6pt plus 2pt minus 1pt}}
}{
  \KOMAoptions{parskip=half}}
\makeatother
\usepackage{xcolor}
\IfFileExists{xurl.sty}{\usepackage{xurl}}{} 
\IfFileExists{bookmark.sty}{\usepackage{bookmark}}{\usepackage{hyperref}}
\hypersetup{
  hidelinks,
  pdfcreator={LaTeX via pandoc}}
\usepackage{graphicx}
\makeatletter
\def\maxwidth{\ifdim\Gin@nat@width>\linewidth\linewidth\else\Gin@nat@width\fi}
\def\maxheight{\ifdim\Gin@nat@height>\textheight\textheight\else\Gin@nat@height\fi}
\makeatother
\setkeys{Gin}{width=\maxwidth,height=\maxheight,keepaspectratio}
\makeatletter
\def\fps@figure{htbp}
\makeatother
\title{An Approach to Causal Inference over Stochastic Networks}

\author[Duncan A. Clark]{Duncan A. Clark}
\address{University of California - Los Angeles, Los Angeles, USA}
\email{duncanclark@ucla.edu}

\author[Mark S. Handcock]{Mark S. Handcock}
\address{University of California - Los Angeles, Los Angeles, USA}

\usepackage{amsmath}\usepackage{amsfonts}\usepackage{amssymb}\usepackage{mathrsfs}\usepackage{graphicx}\usepackage{float}\usepackage{url}\usepackage{epstopdf}\usepackage{booktabs}\usepackage{longtable}\usepackage{appendix}\usepackage[a4paper]{geometry}\usepackage{graphicx}\usepackage[textwidth=8em,textsize=small]{todonotes}\usepackage{amsmath}\usepackage{natbib}\usepackage{hyperref}\usepackage[ruled,vlined]{algorithm2e}\usepackage{xcolor}\usepackage{colortbl}\usepackage{bigdelim}\usepackage{breqn}
\usepackage{booktabs}
\usepackage{longtable}
\usepackage{array}
\usepackage{multirow}
\usepackage{wrapfig}
\usepackage{float}
\usepackage{colortbl}
\usepackage{pdflscape}
\usepackage{tabu}
\usepackage{threeparttable}
\usepackage{threeparttablex}
\usepackage[normalem]{ulem}
\usepackage{makecell}
\usepackage{xcolor}
\ifLuaTeX
  \usepackage{selnolig}  
\fi

\begin{document}

\newtheorem{assumption}{Assumption}
\newtheorem{definition}{Definition}
\newcommand{\R}{\mathbb{R}}
\newcommand{\N}{\mathbb{N}}
\newcommand{\E}{\mathbb{E}}
\newcommand{\V}{\mathbb{V}}
\newcommand{\bfR}{\mathbf{R}}
\newcommand{\bfX}{\mathbf{X}}
\newcommand{\bfW}{\mathbf{W}}
\newcommand{\bfD}{\mathbf{D}}
\newcommand{\INT}{\int_{-\infty}^{+\infty}}
\newcommand{\p}{\partial}
\newcommand{\ra}{\Rightarrow}
\newcommand{\dH}{d\mathscr{H}}
\newcommand{\ch}{\text{cosh}}
\newcommand{\sh}{\text{sinh}}
\newcommand{\ex}{\mathbb{E}\left[X\right]}
\newcommand{\ey}{\mathbb{E}\left[Y\right]}
\newcommand{\indep}{\perp \!\!\! \perp }

\setcounter{secnumdepth}{4}

\begin{abstract}
Claiming causal inferences in network settings necessitates careful consideration of the often complex dependency between outcomes for actors. Of particular importance are treatment spillover or outcome interference effects. We consider causal inference when the actors are connected via an underlying network structure. Our key contribution is a model for causality when the underlying network is endogenous and co-evolves with the actor covariates stochastically over time. We develop a joint model for the relational and covariate generating process that avoids restrictive separability and deterministic network assumptions as these rarely hold realistic social network settings. Our framework utilizes the highly general class of Exponential-family Random Network models (ERNM) of which Markov Random Fields (MRF) and Exponential-family Random Graph models (ERGM) are special cases. We present potential outcome based inference within a Bayesian framework, and propose a modification to the exchange algorithm to allow for sampling from ERNM posteriors. We present results of a simulation study demonstrating the validity of the approach. Finally, we demonstrate the value of the framework in a case-study of smoking over time in the context of adolescent friendship networks.
\end{abstract}
\keywords{Causality, Social Networks, Network models, Spillover, Contagion, Interference, Gibbs measures}

\section{Introduction}\label{sec:intro}

Causal inference is difficult, especially in systems with partially
known and likely complex structure. There is an extensive literature on
causal inference methods for so called ``network settings'' from a
variety of perspectives
\citep{hudgens2008,shalizi2012,ogburn2014,vanderLaan2014,sofrygin2016,deamour2016,aronow2017,shpitser2017,tchetgen2020, ogburn2020_causal}.

There are recent empirical studies claiming strong causal results in
network settings that have been controversial. For example, claims about
the spread of characteristics through social settins, so called social
contagion
\citep{Christakis2008_obese, Christakis2008_dynamics, Christakis2010_sensors}
with corresponding methodological criticism from others
\citep[e.g.][]{ogburn2020_TMLE}. Such results are controversial due not
only to their surprising and perhaps provocative substantive nature, for
example, statements such as ``obesity is socially contagious'', but also
due to the strong assumptions required to justify the methodology.

Much of the problem stems from the unknown underlying social processes.
For example, as explicitly noted in \cite{shalizi2012}, contagion is
often confounded with homophily. They consider Directed Acyclic Graphs
(DAGs) \citep{pearl1995, spirtes2000} and demonstrate that contagion can
be confounded when latent homophily is present.

In this paper we consider social structures represented by networks with
stochastic links and covariates that stochastically co-evolve over time.
We present a generalized chain graph approximation to a credible social
process DAG, which allows for a dependence structure that we believe to
be compatible with such problems. We seek causal inference, that is the
effect on outcomes of a hypothetical intervention. We then frame causal
inference in terms of network equilibrium potential outcomes, that is,
potential outcomes derived from the chain graph structure, and use an
augmented variable Markov Chain Monte Carlo (MCMC) algorithm to sample
from their posterior distributions. The key contribution of our approach
is to allow for uncertainty in the network structure and for
codependence of edges and nodal covariates in the underlying social
process over time.

Chain graphs have been posited as a possible representation of the
causal structure of networks evolving over time that reach equilibrium
\citep{tchetgen2020,shpitser2017}. By considering a DAG of a network
generating process over time, \cite{ogburn2020_causal} and
\cite{lauritzen2002} suggest that estimating causal effects in this
setting is not viable unless fine grained temporal data is available and
the causal structure is simple.

We consider the situation where we observe a single network, considered
a realization of a random social process over both nodes and edges.
Causal inference in this setting cannot be reduced to an independent and
identically distributed (IID) problem in the same fashion as
\cite{sofrygin2016}. The chain graph approximation employed in
\cite{tchetgen2020} and \cite{ogburn2020_causal} cannot be used as it
does not allow for stochastic connections. There have been efforts to
allow for network uncertainty \citep{toulis2013, Kao2017}. However these
methods require unconfoundedness of the edges and the random nodal
covariates, given enough fixed information about the nodes. In not
making such an assumption we require a joint probability model for the
random edges and nodal covariates in the graph.

We note that the literature of network causal inference is at present
sharply disconnected from the literature concerning generative models
for social networks. As most approaches have considered the networks as
fixed, there has been little interest in placing a probability
distribution over the space of possible networks. Typically generative
models for social networks for example the commonly used
Exponential-family Random Graph Models (ERGM) \citep{FrankStrauss1986}
consider the edges as the random variables to be modelled and nodal
covariates as fixed. There has been much sophisticated work on
understanding such models
\citep{handcock2003, Robins2007, schweinbergerhandcock2015}, and well
developed MCMC based fitting procedures \citep{Snijders2002, han02} with
associated complex software \citep{ergm2008}. However due to the assumed
fixed nature of the nodal covariates, these are of limited use for
causal inference on nodal outcomes. Markov random field (MRF) models
treat the nodal covariates as stochastic but the connections between
nodes fixed. Encompassing both model classes are the novel
Exponential-family Random Network Model (ERNM) \citep{fellows2012}. ERNM
are a class of exponential family models that encompasses ERGMs and MRF
as special cases. ERNM allows for the edges and nodal covariates to be
stochastic, thus the nodal covariates and the edges can be co-dependent.
The focus of this paper is causal inference based on the plausible
representation of complex social structure via ERNM.

We utilize a Bayesian framework for the causal quantities and the
network model. This allows for the incorporation of prior information,
as well the automatic accounting of uncertainty in a theoretically
consistent manner. The Bayesian approach does not require appeals to
asymptotic arguments for its validity. Indeed, asymptotics for causal
quantities in our setting are conceptually difficult as there is no
single asymptotic framework that is compelling. In particular, the
number of nodes, \(N\), is a fundamental characteristic of the social
process and not a sampling design characteristic, as it is in most of
Statistics. For example, the interactions of a class with 5 students
will be quite different from a class of size 75. Hence asymptotic
approximations must identify credible invariant parametrisations
\citep{krivitsky2011}. Different values of \(N\) change the fundamental
structure of the social network, as dependent edge behaviour is strongly
related to the number of nodes in a network. We develop a modification
to the exchange algorithm \citep{Murray2006} to allow for sampling from
ERNM posteriors, which we use to infer the posterior distributions of
potential outcomes and, hence, estimate causal estimands of interest.

This paper is structured as follows. Section \ref{sec:setup} introduces
our general network setting and our notation. Section
\ref{sec:causal_framework} considers the DAG of a network process over
time allowing for network uncertainty with a chain graph approximation.
Section \ref{sec:potential_outcomes} defines causal quantities of
interest in terms of equilibrium potential outcomes. Section
\ref{sec:sims} briefly describes a simulation study, the details of
which are contained in the supplement. Section \ref{sec:example}
considers a case-study of a network from the National Longitudinal Study
of Adolescent Health \citep{AddHealth2007} and gives estimates of
unknown causal quantities. Section \ref{sec:discussion} provides general
discussion of the method, and its ability to generate credible causal
inference.

\section{Notation and Setting}\label{sec:setup}

We consider a known fixed set of \(N\) nodes. Each node has a random
nodal outcome \(Y_{i}\), thus the whole network outcome is
\(Y = (Y_{1},\ldots Y_{N})\). Realizations of the random nodal
covariates \(Y_{i}\) and \(Y\) are denoted with lower case \(y_{i}\) and
\(y\). For this paper we only consider binary outcomes (although the
ideas are easily extended to non-binary outcomes). Each node is also
permitted to have further multivariate nodal covariates similarly
denoted \(X = (X_{1},\ldots X_{N})\) with \(X_{i} \in \mathbb{R}^{p}\)
for some p.

We denote the random edges between nodes as the random variable
\(A = \lbrace A_{i,j} \rbrace_{i=1,j=1}^{N}\), with realizations \(a\).
\(A\) can be considered a random adjacency matrix. We also restrict
\(A_{i,j}\) to be binary with \(1\) indicating a connection and \(0\)
representing the absence of a connection. For this paper, we make the
restriction that our networks are undirected
i.e.~\(A_{i,j} = A_{j,i}\quad \forall i,j\).

A network realization is defined to be a set \(\lbrace y,a,x \rbrace\).
When considering the dynamic network of the process over time we
indicate the outcomes at time \(t\) with superscript, for example, the
outcome random variable for node \(i\) at time \(t\) is \(Y_{i}^{t}\)
and the whole network random variable at time t is
\(\lbrace Y^{t},A^{t},X^{t} \rbrace\). If the temporal dynamics result
in an equilibrium distribution, we denote it by
\(\underset{t\rightarrow\infty}{\lim} \left( Y^{t},A^{t},X^{t}\right) = \left(Y,A,X \right)\).

As the node set is fixed, the nodal covariates \(X\) are often in
practice fixed throughout the evolution of the social process. Going
forward we omit \(X\) from our notation, that is for clarity, we
consider our networks as realizations of
\(\left(Y,A\right) \vert X = x_{observed}\) but write
\(\left(Y,A\right)\).

We represent the treatment of nodes via the treatment vector
\(Z = (z_1, \ldots z_N)\) with realizations \(z\). For the purposes of
this paper we consider the treatment to be applied prior to the
evolution of the network process though perhaps conditional on the fixed
nodal covariates. We leave allowing for time varying treatments
assignments and outcome evolution to future research, though we believe
it to be compatible with our approach.

In the following section we will introduce a DAG to represent the
dependence structure of the social network. We emphasize that the nodes
in the DAGs represent random variables in the stochastic social process
underlying the above described social network setting (rather than
actors in the social network). That is, nodes in the DAG represent
random variables in the social process generating the network, they may
either be treatments, random nodal covariates, or random edges.

\section{Network Potential Outcomes and Causal Estimands of Interest}\label{sec:potential_outcomes}

\cite{hudgens2008}, \cite{toulis2013} and \cite{aronow2017} all
considered potential outcome-based frameworks of assumptions and
definitions as a basis for causal inference for nodal covariates. We
consider network potential outcomes as realizations of an equilibrium
distribution of a social process that evolved over time.

Our causal estimands should be interpreted as the effect of an
intervention on the equilibrium distribution, this is implicit in
\cite{toulis2013} and \cite{aronow2017}. Estimands in
\cite{tchetgen2020} and \cite{ogburn2020_causal} are based interventions
on nodal statuses prior to the evolution of the social process, and
estimate network effects, rather than nodal effects. We note that these
whole network direct or spillover treatment outcomes for pre social
processes interventions, the ERNM \citep{fellows2012} presented in
Section \ref{sec:modelling} is still fully compatible with such regimes,
and indeed a generalization of the MRF models used in
\cite{tchetgen2020} and \cite{ogburn2020_causal}.

We set \(Y_{i}(Z,A,Y_{-i})\) to be the equilibrium potential outcome of
node \(i\), given the treatment vector of all nodes \(Z\), the network
\(A\), and the outcomes of all the other nodes \(Y_{-i}\). In this
definition, each potential outcome \(i\) depends fully on the entire
rest of the edges and nodal potential outcomes in the network. However
as our estimands of interest relate to one-step neighbourhoods of nodes,
we make the following assumption.

The one-step neighbourhood, assumptions also allows us to dramatically
reduce the required number of simulations to estimate missing potential
outcomes in section \ref{sec:estimation}.

\begin{assumption}{One Step Neighbourhoods}\label{assump:neighbourhood_treatment}
\begin{align}
\forall i \qquad Y_{i}(Z,A,Y_{-i}) = Y_{i}(Z_{i},Z_{\mathscr{N}_{i}(A)},\mathscr{N}_{i}(A),Y_{\mathscr{N}_{i}(A)})
\end{align}

{\noindent}where $\mathscr{N}_{i}(A)$ is the neighbourhood of node $i$ in the network $A$. That is, the individual outcomes only depend on treatments, edges and outcomes in their own one-step neighbourhood.
\end{assumption}

We next state the causal estimands that we will pursue in their full
generality. In section \ref{sec:estimation} we will consider models, and
simplifying assumptions that allow for inference in practice.

\begin{definition}{Primary Effects}
\begin{align}
\xi_{i} &= \frac{1}{|(Z^{-i},A,Y^{-i})|} \sum_{(z^{-i},a,y^{-i}) \in (Z^{-i},A,Y^{-i}) } Y_{i}(z_{i} = 1, z,a,y) - Y_{i}(z_{i} = 0, z,a,y) \\
\xi^{ave} &= \frac{1}{N}\sum_{i=1}^{N} \xi_{i}
\end{align}

{\noindent}where $(Z^{-i},A,Y)$ are all possible combinations of edges, outcomes and treatments excluding the outcome and treatment of node i.

\end{definition}

\begin{definition}{$k$-peer Treatment Effect}
\begin{align}
\delta^{k}_{i} &= \frac{1}{|(Z,A)_{\mathscr{N}_{i}}|} \sum_{(z,a,y) \in (Z,A)^{k}_{\mathscr{N}_{i}} } Y_{i}(z_{i} = 0, z,a,y) -  Y_{i}(z_{i} = 0, (z,a) \in (Z,A)^{0}_{\mathscr{N}_{i}},y)\\
\delta_{k} &= \frac{1}{N}\sum_{i = 1}^{N} \delta^{k}_{i}
\end{align}

{\noindent}where $(Z,A)^{k}_{\mathscr{N}_{i}}$ is the set of combinations of node-i-local edge realizations and treatment assignments such that node $i$ is exposed to $k$ other treated nodes.

\end{definition}

\begin{definition}{$k$-peer Outcome Effect For Binary Outcomes}\label{def:k_peer_out}
\begin{align}
\delta^{k}_{i} &= \frac{1}{|(A,Y)_{\mathscr{N}_{i}}|}  \sum_{(z,a,y) \in (A,Y)^{k}_{\mathscr{N}_{i}} } Y_{i}(z_{i} = 0, z,a,y) -  Y_{i}(z_{i} = 0,z, (a,y) \in (A,Y)^{0}_{\mathscr{N}_{i}})\\
\delta_{k} &= \frac{1}{N}\sum_{i = 1}^{N} \delta^{k}_{i} 
\end{align}

{\noindent}where $(A,Y)^{k}_{\mathscr{N}_{i}}$ is the set of combinations of node-i-local edge and outcome realizations such that node $i$ is exposed to $k$ other nodes with outcome $1$.
\end{definition}

We note that the \(k\)-peer outcome effect could be considered as a
special case of a \(k\)-peer treatment effect where the treatment
results in outcome \(1\), almost surely. However we find it convenient
to express the estimand separately as it is often of substantive
interest in networks where some other treatment is administered, but the
\(k\)-peer outcome effect is important.

\section{Causal Framework}\label{sec:causal_framework}

DAGs provide a means for precisely specifying the structure of
relationships between random variables (see \cite{pearl2009} for an
introduction). \cite{pearl1995} developed strict criteria for
identification of causal effects. Formal equivalence with the potential
outcome framework was shown in \cite{richardson2013}

Nodes in the DAG represent random variables, with edges drawn between
nodes being strictly uni-directional, and cycles of edges prohibited, so
that so called ``feedback'' loops are not permitted. The interpretation
of a directed line from node \(i\) to node \(j\) is that \(i\) causally
effects \(j\). For a node indexing set \(V\), denote the corresponding
random variables \(\lbrace x_{v} \rbrace_{v\in V}\). This graphical
structure encodes the following factorization of their joint probability
distribution \(p(x) = \prod_{v \in V} p(x_{v} \vert x_{pa(v)})\), where
\(pa(v)\) are the parents of \(v,\) that is the nodes in the DAG with
edges into \(v\).

The causal effect of \(X\) on \(Y\) is represented by
\(p(y\vert do(X=x)),\) the distribution of the random variable \(Y\)
when \(X\) has been externally set to \(x\). \cite{pearl2009} provides
transformations for expressing this new distribution
\(p(y\vert do(X=x))\) in terms distribution on observed random variables
e.g.~\(p(y\vert x)\) or \(p(y\vert x,z)\) for \(z\) some other variable
in the system.

Chain graphs permit undirected as well as directed edges, which allow
for different Markov properties from DAGs. In fact chain graphs can
represent dependence structures that are not possible under a DAG. We
include some discussion in the supplement however we omit subtleties of
their Markov properties discussed in \cite{frydenberg1990} and utilized
for causal analysis in \cite{lauritzen2002}. In their full generality,
chain graphs can express complex dependence structures. However, in our
case, our example chain graph only has one chain component which results
in none of the outcomes being rendered conditionally independent, thus
we do not require an in-depth review for the purposes of this paper.
Practically, one possible interpretation of the undirected edges in a
chain graph is that the two variables interact in a causal feedback
sense over time.

DAGs representing the causal structure of outcomes of nodes in networks
under both interference and contagion are given a clear and detailed
treatment in \cite{ogburn2014}. We generalize the related conjecture in
\cite{ogburn2020_causal} for a chain graph approximation of causal
structure of a social network, which slowly evolves over time. This is
based on an interpretation of causality with feedback relationships over
time which chain graphs can be used to explain \citep{lauritzen2002}.
Our generalization is to consider a social network where the connections
are not fixed and are motivated by the empirical observation that social
connections are often strongly dependent on other nodes' connections as
well as other nodes' covariates, treatments and outcomes.

\begin{figure}[!ht]
\begin{center}
\includegraphics[width=\textwidth]{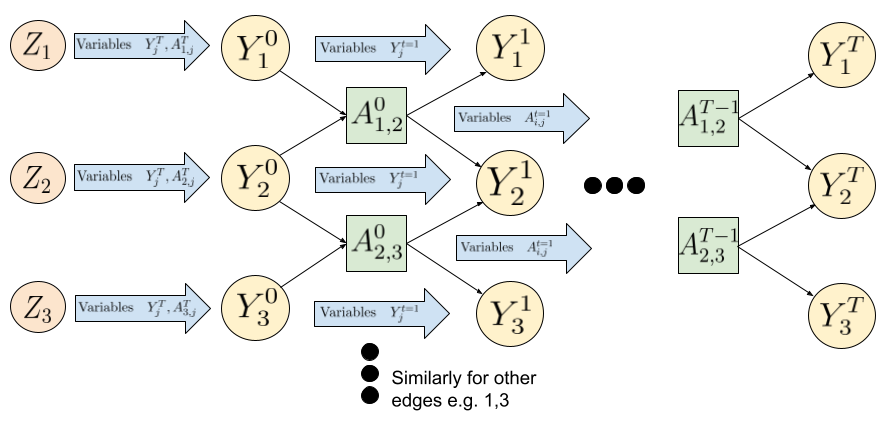}
\caption{Full DAG of temporal network formation feedback process, in the three node network case. Treatment variables $Z_j$ are permitted to causally effect all of the other variables. Outcomes $Y_{i}^{t}$ and $Y_{j}^{t}$ are permitted to causally effect edges $A^{t}_{i,j}$, as well as outcomes $Y_{i}^{t+1}$ and $Y_{j}^{t+1}$. Edges $A_{i,j}^{t}$ are permitted to causally effect edges $A_{k,l}^{t+1}$ with no restrictions on $k$ and $l$. In practice the functional form of models will usually restrict the causal impact of node $i$'s outcome and treatment to only the set of neighbouring nodes $\lbrace j \rbrace$, that is, where there is an edge present between node $i$ and $j$.}
\label{fig:full_DAG}
\end{center}
\end{figure}

As an illustration for our chain graph approximation, Figure
\ref{fig:full_DAG} represents the structure of a network evolving over
time for a three node network similar to that of
\cite{ogburn2020_causal} and \cite{tchetgen2020}. Note the analogous
connections between node 3 and 1 are omitted for clarity. Block arrows
denote multiple arrows into the variables described therein. The full
DAG with all arrows becomes quickly unwieldy. In the notation of Section
\ref{sec:setup}, the nodal treatments are
\(\lbrace Z_{1}, Z_{2}, Z_{3} \rbrace\), nodal outcome variables
\(\lbrace Y_{1}, Y_{2}, Y_{3} \rbrace\), and we denote the undirected
edges between nodes \(i\) and \(j\) as \(A_{i,j}\). Directed network
edges as well as fixed nodal covariates can also be built-in but are
omitted here for clarity. Superscripts denote the status of a variable
at that time step e.g.~\(Y_{2}^{t}\) is the outcome for node \(2\) at
time \(t\).

Under this DAG, each of the treatment variables \(Z_j\) are permitted to
causally effect all of the other outcome variables, as well as edges
involving \(j\). Outcomes \(Y_{i}^{t}\) and \(Y_{j}^{t}\) are permitted
to causally effect edges \(A^{t}_{i,j}\), as well as outcomes
\(Y_{i}^{t+1}\) and \(Y_{j}^{t+1}\). Edges \(A_{i,j}^{t}\) are permitted
to causally effect edges \(A_{k,l}^{t+1}\) with no restrictions on \(k\)
and \(l\).

We note that there are key edges that are not present in this DAG. Edges
and outcomes, from time step \(t\) can only influence edges and outcomes
at time step \(t+1\), not others. We argue that this is plausible under
slow evolution of a network process over time. In general, it is
exceedingly rare that enough data are available to identify all the
relationships posited in Figure \ref{fig:full_DAG}.

The DAG for our system is highly complex, and we usually observe little
incremental time data on the social process. Thus operations as
introduced in \cite{pearl1995} to reduce \(p(y\vert do(X=x))\) to
expressions involving only distributions, from which we have
realizations, e.g., \(p(y\vert x)\), are not possible. That is the true
causal effect is unidentifiable. We pursue ``approximate causal
inference'' through approximating the DAG for our social process over
time with a chain graph.

\begin{figure}[!ht]
\begin{center}
\includegraphics[width=\textwidth]{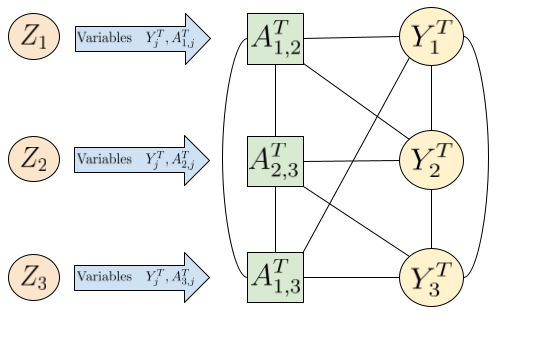}
\caption{Chain graph approximation of a three node temporal social network process with treatment}
\label{fig:3_node_chain}
\end{center}
\end{figure}

As noted in \cite{lauritzen2002}, the equilibrium distribution of a so
called infinite DAG can be represented as a chain graph. Similarly to
\cite{ogburn2020_causal}, but modelling edges as random, the DAG in
Figure \ref{fig:full_DAG} can be approximated by the chain graph shown
in Figure \ref{fig:3_node_chain}. The chain component (undirected
component) of this chain graph is close to being complete since we allow
every nodal outcome to influence every other nodal outcome, however we
only allow outcomes \(Y_{i}\) and \(Y_{j}\) to influence edge
\(A_{i,j}\). In the full DAG, backdoor paths through previous time steps
result in this not being the case.

This suggests that the complex structure of such a DAG can be
approximated with the chain graph factorization:

\begin{align}\label{eq:chain_factorization}
p(Y,A,Z) = \left( \prod_{i=1}^{n} p(z_{i})\right) p(Y,A \vert Z)
\end{align}

As an approximation to the true temporal DAG a chain graph model of
causality serves to render the causal effects tractable in practice. See
\cite{ogburn2020_causal} and \cite{lauritzen2002} for a fuller
explanation.

As an illustration of the generality of this chain graph's dependence
structure we define a conditional independence property, that we refer
to as \textsl{local conditional independence} as follows:

\begin{align}\label{eq:local_network_outcome}
Y_{i}^{T} \perp \!\!\! \perp Y_{j}^{T}, Z_{j}~~ \Bigg\vert ~~A_{i,j}^{T} = 0,~~Z_{i}, ~~\lbrace Y_{l}^{T} : A_{i,l}^{T} = 1 \rbrace
\end{align}

That is, nodal outcomes are conditionally independent given all
neighbours, and that they are not connected. Equation
(\ref{eq:local_network_outcome}) does not hold \textit{a priori} due to
the dependence induced by the random edge \(A_{i,j}\) unlike in the
fixed network case where it does. That is, the chain graph retains
dependencies between outcomes of non-connected nodes, even when
conditioning on neighbours. In fact, due to the close-to-complete nature
of the chain component, there are few conditional independence
assumptions that can be concretely made. Indeed the complexity of such
systems is the reason why social network modelling has proven to be
difficult.

Conditional independence properties have been considered for ERGMs
\citep{snijders2006}. Common properties induced by model choice in ERGMs
are the so-called \textit{Markov} property \citep{FrankStrauss1986} and
the so called \textit{social circuit} property that requires the Markov
property in addition to a cycle condition \citep{snijders2006}. Both of
these commonly imposed assumptions severely limit possible dependence
structures. In practice the social network modelling community has found
the social circuit assumption to yield well fitting models in a wide
variety of situations \citep{Goldenberg2010}.

We do not comment on the validity of the approximation of the DAG by the
chain graph. \cite{ogburn2020_causal} gave simulations supporting their
version of this approximation. We will only consider recovering causal
estimands based on the assumption that the chain graph approximation
holds.

The remainder of this paper concerns modelling \(P(Y,A\vert Z)\) in
order to estimate causal quantities. We specify our causal estimands
using equilibrium distribution potential outcomes. Implicitly the
assumption we make in the DAG formulation, under some intervention on
the equilibrium denoted with a superscript \(*\).

\begin{align}
P_{equil}(Y_{i} = y \vert (A,Z,Y_{i}^{c}) = (A,Z,Y_{i}^{c})^{\ast}) = P_{equil}(Y_{i} = y \vert \text{do}( (A,Z,Y_{i}^{c}) = (A,Z,Y_{i}^{c})^{\ast})
\end{align}

We note that intervening on an equilibrium distribution is not possible
in practice. However in order to claim the estimation of causal
quantities, which are intertwined with social processes, we believe it
to be necessary as well as implicit in current potential outcome based
approaches \citep{aronow2017,toulis2013}. We interpret our causal
estimands as an average, of the possible treatments, and social
processes that led to a given treatment in equilibrium.

The lack of conditional independence assumptions that we are able to
make in Section \ref{sec:estimation}, is the direct consequence of the
nearly complete chain graph model specified in this section.

\section{Estimation and Identifying Assumptions}\label{sec:estimation}

Section \ref{sec:potential_outcomes} introduced network potential
outcomes and our estimands of interest. Section
\ref{sec:causal_framework} justified that they have a casual
interpretation. In this section we will introduce models that can be
practically used to infer the missing potential outcomes, to produce
concrete estimates of causal effects.

Our strategy is to pursue model-based Bayesian imputation of potential
outcomes \citep{imbens2015}.

\subsection{Structural Assumptions}

We next state possible assumptions restricting the dependence structure
of the process that will enable us to feasibly impute missing potential
outcomes from network simulations. In our assumptions we explicitly
include fixed nodal covariates \(X\).

\begin{assumption}{Unconfounded Treatment Assignment Assumption Under Network Assignment}\label{assump:treat_assump}
\begin{align*}
P(Z\vert X,A,\mathbb{Y}) = P(Z\vert X,A,\mathbb{Y}^{\prime}) \quad \forall Z,X,A, \mathbb{Y}\quad and\quad \mathbb{Y}^{\prime}
\end{align*}
{\noindent}where $A$ is a network, X is a set of nodal covariates on the nodes, and $\mathbb{Y}$ are the potential outcomes.
\end{assumption}

To account for uncertainty in the network it is possible to make the
assumption, as in \cite{Kao2017}, that the causal link between the
network and outcomes can be broken by the inclusion of nodal covariates.
Specifically:

\begin{assumption}{Unconfounded Network Assumption under Network Interference}\label{assump:net_assump}

\begin{align*}
P(A\vert X ,\mathbb{Y}) = P(A\vert X, \mathbb{Y}^{\prime}) \quad \forall A,X,\mathbb{Y}\quad and\quad \mathbb{Y}^{\prime}
\end{align*}

\end{assumption}

We note that Assumption \ref{assump:treat_assump} also follows as a
consequence of the chain graph approximation, as in the three node
example in Figure \ref{fig:3_node_chain}
\citep[See][for discussion on this]{lauritzen2002}.

Assumption \ref{assump:net_assump} essentially assumes away the main
problem of network causal inference, that the network structure and the
potential outcomes are related. The idea is that after including enough
nodal covariates, the association between the network and the nodal
potential outcomes breaks down. Note that this assumption trivially
holds if the network is considered fixed.

Denoting the missing potential outcomes as \(\mathbb{Y}_{miss}\) and the
observed as \(\mathbb{Y}_{obs}\). Using Assumptions
\ref{assump:neighbourhood_treatment} and \ref{assump:treat_assump},
\cite{Kao2017} show that : \begin{align}\label{cond_indep_1}
P(\mathbb{Y}_{miss} \vert X, Z, A, \mathbb{Y}_{obs}) = P(\mathbb{Y}_{miss} \vert X, \mathbb{Y}_{obs})
\end{align}

\cite{Kao2017} propose choosing covariates \(X\) to include so that
Assumption \ref{assump:net_assump} is met, and suggest that the addition
of covariates derived from modelling the network \(A\) may be included
to achieve this.

We require only Assumptions \ref{assump:neighbourhood_treatment} and
\ref{assump:treat_assump}, and modelling the potential outcomes jointly
with the network. We argue that this is more realistic in most
situations, and forces the researches into more realistic modelling
choices to arrive at causal inference for nodal outcomes.

The relaxation of Assumption \ref{assump:net_assump} breaks down the
proof of Equation (\ref{cond_indep_1}), with the analogous result under
the relaxed assumption being: \begin{align}
P(\mathbb{Y}_{miss} \vert X, Z, A, \mathbb{Y}_{obs}) = P(\mathbb{Y}_{miss} \vert X, A, Y_{obs})
\end{align}

The retention of the network \(A\) in the conditioning variables,
requires that we model its posterior distribution in the imputation of
the missing potential outcomes.

\subsection{Modelling}\label{sec:modelling}

To facilitate modelling when the network is not fixed, we use the ERNM
\citep{fellows2012}. The ERNM model can be viewed as a generalization of
ERGM to allow for random nodal covariates, alternatively and
equivalently it can be viewed as generalization of Markov Random Fields
(MRF) that allows for random edges. The basic formulation is an
exponential family for network \(a\) with nodal covariates \(y\):

\begin{align}
P(A=a, Y=y \vert \theta) = \frac{1}{c(\theta,\mathscr{N})} \exp\left(\theta \cdot g(a,y) \right)
\end{align}

The sample space \(\mathscr{N}\) is the space of all possible binary
edge realizations together with all of the possible random nodes, e.g.,
\(\mathscr{N} \subset 2^{\mathbb{A}} \times \mathscr{Y}^{n}\), where
\(2^{\mathbb{Y}}\) is the power set of all the dyads and
\(\mathscr{X}^n\) is the joint sample space of the nodal covariates. For
formal details see \cite{fellows2012}.

MRF models can be seen as the ERNM conditional on the network in
Equation (\ref{eq:ERNM_to_MRF}), and ERGMs the ERNM conditional on the
nodal attributes. The normalising for the conditional distribution
constant for the MRF is written \(c(\theta,\mathscr{N}(a),a)\) to
reflect summing over the restricted space \citep{fellows2012}.

\begin{align}\label{eq:ERNM_to_MRF}
P(Y=y \vert A = a, \theta) =  \frac{1}{c(\theta,\mathscr{N}(a),a)} \exp\left(\theta \cdot g(a,y) \right)
\end{align}

For example a typical MRF model on a network with treatment effect,
outcome homophily as well as outcome and treatment neighbour effects can
be written as:

\begin{align}\label{eq:MRF_example}
p(Y = y \vert A=a, Z=z) = \frac{1}{c(\theta,\mathscr{N}(a),a)}\exp \big(
& \theta_{1} \cdot \sum_{i=1}^{N} y_{i} +
\theta_{2} \cdot \sum_{i=1}^{N} y_{i}z_{i} + \\
&\theta_{3} \cdot \sum_{i,j =1, a_{i,j} =1}^{N} \mathbb{I}(y_{i}=y_{j}) + \\
& \theta_{4} \cdot \sum_{i,j =1, a_{i,j} =1}^{N} y_{i}Y_{j}+
\theta_{5} \cdot \sum_{i,j =1, a_{i,j} =1}^{N} y_{i}z_{j}\big)
\end{align}

The full ERNM, however, is permitted to include terms typically included
to account for common social phenomenon, e.g.~social transitivity - the
tendency for edges to complete triangles of edges, or social popularity
- the tendency for some nodes to have many more connection than others.
For example the number of edges, triangles, two-stars and three-stars
may used. A typical ERNM model which in addition the the MRF terms in
equation \ref{eq:MRF_example} accounts for transitivity, and
centralisation with triangles and two-stars can be written as:

\begin{align}\label{eq:ernm_example}
p(Y = y \vert A=a, Z=z) = \frac{1}{c(\theta,\mathscr{N}(a),a)}\exp \big(
&\theta_{1} \cdot \sum_{i=1,j=1}^{N} a_{i,j} + \\
&\theta_{2} \cdot \sum_{i,j,k=1}^{N} \mathbb{I}(a_{i,j} = a_{i,k} = a_{j,k} = 1) + \\
&\theta_{3} \cdot \sum_{i,j,k=1}^{N} \mathbb{I}(a_{i,j} = a_{i,k} = 1) +
\theta_{4} \cdot \sum_{i=1}^{N} y_{i} + \\
&\theta_{5} \cdot \sum_{i=1}^{N} y_{i}z_{i} +
\theta_{6} \cdot \sum_{i,j =1, a_{i,j} =1}^{N} \mathbb{I}(y_{i}=y_{j})+ \\
&\theta_{7} \cdot \sum_{i,j =1, a_{i,j} =1}^{N} y_{i}y_{j}+
\theta_{8} \cdot \sum_{i,j =1, a_{i,j} =1}^{N} Y_{i}z_{j}\big)
\end{align}

\cite{tchetgen2020} utilized Markov random field (MRF) models with
coding or pseudo likelihood estimators to power a Gibbs sampling
procedure, from which they estimated causal effects of treatment in
situations where a single network was observed and the observed data
treatment was permitted to be affected by covariates. It is not clear
why modern MCMC methods were not used, as pseudo likelihood methods are
known to have undesirable properties \citep{vanDuijn2009}.
\cite{ogburn2020_causal} gave an example with parameters estimated from
multiple observations from a MRF using the to ``maximize penalized
node-conditional likelihoods''.

\cite{fellows2012} extended extensive work on MCMC MLE estimation for
ERGM \citep{Snijders2002,Hunter2006} to ERNM. However following the
Bayesian paradigm and we simulate from the posterior distribution of the
missing potential outcomes conditional on the observed data. This
accounts for uncertainty in a theoretically consistent manner
\cite{ohagan1993}, and removes the need for asymptotic assumptions on
the node set which are unrealistic, or bootstrapping \cite{efron1979} as
in \cite{ogburn2020_causal}.

\cite{toulis2013} also followed this paradigm, allowing for edge
uncertainty with a Poisson edge model, and a linear outcome model. We
note that their model does not account for the dyad dependent nature of
real social processes, which is perhaps the most important feature of
social network data.

We note that sampling from the posterior distribution of an ERNM is
non-trivial, details are contained in the supplement, which require the
use of a the exchange algorithm for so called doubly intractable
distributions \citep{Murray2006}.

With a suitably simple MRF model with a fixed network or a separable
network and outcome model, the causal effects can usually be computed
directly from the realized parameter values. For an ERNM this is not the
case. Noting that, in full generality, each of the nodal potential
outcomes depends on the whole network and all other nodal potential
outcomes. The equilibrium distribution of a missing binary potential
outcome for node \(i\) can be written:

\begin{align}
P(\mathbb{Y}_{miss,i}(a,y^{-i})\vert X, A_{obs}, Y_{obs}) &= \int_{\Theta} P(\mathbb{Y}_{miss,i}(a,y^{-i})\vert \theta , X, A_{obs}, Y_{obs}) p(\theta \vert X, A_{obs}, Y_{obs}) d\theta
\end{align}

We can then approximate this by simulating a large number of networks
\(\lbrace (A,Y)_{j} \rbrace_{j=1}^{M}\), letting
\(M(a,y^{-i}) = \lbrace (a^{\prime},y^{\prime}) \in \lbrace (A,Y)_{j} \rbrace_{j=1}^{M} : a^{\prime} = a, y^{\prime,-i} = y^{-i}\rbrace\)
be the number of these networks with the required network \(a\) and
other nodal covariates \(y^{-i}\). This yields

\begin{align}\label{eq:full_approx}
P(\mathbb{Y}_{miss,i}(a,y^{-i}) = 1 \vert X, A_{obs}, Y_{obs}) &\approx  \frac{1}{\vert M(a,y^{-i})\vert}\sum_{(a,y) \in M(a,y^{-i})} Y_{miss,i}(a,y)
\end{align}

Simulating enough networks that have the required network and nodal
covariates is infeasible for networks of realistic size. We consider
one-step neighbourhoods, that is we allow only nodes connected to the
ego to effect the nodal outcome. Thus we dramatically reduce the number
of unique potential outcomes for any given node, by requiring that only
the treatment assignment, edges involving a node and outcomes of the
neighbours of the nodes matter, not the whole network.We note that this
is, in fact, a highly restrictive assumption, though is required to
feasibly simulate the missing potential outcomes.

Concretely to estimate the potential outcome for the k peer effect
estimand, for each \(i\), instead of restricting to \(M(a,y^{-i})\) we
restrict to
\(M_{\mathbb{N}_{i}}^{k}(a,y^{-i}) = \lbrace (a^{\prime},y^{\prime}) \in \lbrace (A,Y)_{j} \rbrace_{j=1}^{M} : (a^{\prime},y^{\prime}) \in (A,Y)^{k}_{\mathbb{N}_{i}}\rbrace\).
Where \((A,Y)^{k}_{\mathbb{N}_{i}}\) is defined in Definition
\ref{def:k_peer_out}. That is we only restrict to simulations where the
correct neighbourhood is achieved to estimate the expected value of the
missing potential outcomes.

\begin{align}\label{eq:part_approx}
P(\mathbb{Y}_{miss,i}(a,y^{-i}) = 1 \vert X, A_{obs}, Y_{obs}) &\approx  \frac{1}{\vert M_{\mathbb{N}_{i}}^{k}(a,y^{-i}) \vert}\sum_{(a,y) \in M_{\mathbb{N}_{i}}^{k}(a,y^{-i}) } Y_{miss,i}(a,y)
\end{align}

We can then use these expected potential outcomes, to estimate
(Bayesian) expected versions of the causal estimands conditional on the
observed data.

\section{Example : Simulation Study}\label{sec:sims}

\subsection{A DAG compatible data generating process}

In this section we consider simulating from a \(100\) node network, with
a procedure that follows the true DAG for a social process. Letting
\(A\) now be the edge random variables, and \(Y\) the outcome random
variable. We propose the following simulation procedure. \bigskip

\begin{algorithm}[H]\label{algo:DAG_algo}
\SetAlgoLined
\KwResult{$(A,Y)$ sampled from $2^{\mathbb{A}} \times \mathscr{Y}^{n}$}
 Assign treatments Z \\
 Let $Y_{i} = 0 \quad \forall \quad i$ \\
 Let $A_{i,j} = 0 \quad \forall\quad i,j$ \\
 \For{$t = 1,2, \ldots T$}{
  Simulate $Y^{t}$ from $p(Y^{t} \vert Y^{t-1}, A^{t-1},Z)$\\
  Simulate $A^{t}$ from $p(A^{t} \vert Y^{t},A^{t-1},Z)$\\
 }
\caption{Figure \protect\ref{fig:3_node_chain} DAG simulation procedure}
\end{algorithm}
\bigskip

The algorithm specified in Algorithm \ref{algo:DAG_algo} is deliberately
abstract, we do not specify the probability functions
\(p(Y^{t} \vert Y^{t-1}, A^{t-1},Z)\) or
\(p(A^{t} \vert Y^{t},A^{t-1},Z)\) yet.

For our simulations we suggest that choosing
\(p(Y^{t} \vert Y^{t-1}, A^{t-1},Z)\) and
\(p(A^{t} \vert Y^{t}, A^{t-1},Z)\) as a logistic regression, using
change statistics as predictors. That is we allow for a proposed tie or
node change to be more or less probable based the corresponding change
to some specified network statistics. We suggest choosing change
statistics in line with our intuition on social processes, for example
edges that complete triangles of edges are, all else equal, more likely
to form than other edges.

We make a slight simplification: we only allow a single edge or node to
toggle at each time step. This results in the probability of a step
being exactly the acceptance probability that would be used if we were
using a Markov chain to sample from an ERNM, with a simple \(1\) edge or
\(1\) vertex toggle proposal step. Thus sampling from the DAG with this
kind of model for a large enough \(T\) is equivalent to sampling from a
Markov chain for the corresponding ERNM.

The purpose of implementing simulations is to provide empirical evidence
to convince the reader that our method can recover causal effects in the
real world case where the true data generating process (DGP) is complex
and unknown. The credibility generated by the exercise depends on the
credibility of the chosen DGP. If a simplistic DGP is chosen that
deliberately fits the method well, the simulations generate less
credibility. The fact that an ERNM sampling procedure can well represent
the suggested DAG, should not dissuade the reader of the value of the
simulations, in fact that the DAG is compatible with ERNM suggests that
our model may in fact be less mis-specified than feared.

In particular with the ERNM DGP it is possible to generate networks with
transitivity, homophily and contagion, yet still know the posterior
distribution of the causal estimands, for a given network simulation.
Other generating processes could have been selected, but would limit the
scope of understanding the performance of our method as the true
posterior would not be available.

\subsection{Model Specification}

We consider \(4\) possible DGPs for \(100\) node networks where \(50\%\)
for the nodes are treated before the social process evolves. We simulate
networks from the ERNM DGP then fit the posterior distributions under
that DGP which generates the ground truth posterior distribution. We
then fit the posteriors of the remaining three mis-specified DGPs to the
simulated networks, and compare the resulting posteriors to the ground
truth posteriors.

Table \ref{tab:DGP} shows the proposed parameter values and key
properties of the model classes for the \(4\) DGPs. These ERNM
parameters were chosen for simplicity and to achieve a mean degree of
close to \(3\), which might be reasonable in for example a friendship
network. We only present the results of simulation from the ERNM model,
as it is the only model that represents the DAG.

\begin{table}

\caption{\label{tab:unnamed-chunk-1}\label{tab:DGP}Data Generating Process Summary. First block is parameter values for the edge model, second block is the parameter values for the node model, which in the case of the ERNMs are not separable. The third block gives a basic summary of the model classes. We consider networks generated by the ERNM and fit with the other DGPs, the parameter values for the other DGPs are shown to demonstrate the terms included in those models, not for model fitting.}
\centering
\begin{tabular}[t]{lllll}
\toprule
  & ERNM & MRF & ERGM+Logistic & Logistic\\
\midrule
\cellcolor{gray!6}{Edges} & \cellcolor{gray!6}{4.5} & \cellcolor{gray!6}{NA} & \cellcolor{gray!6}{4.5} & \cellcolor{gray!6}{NA}\\
GWESP & 1 & 1 & 1 & NA\\
\cellcolor{gray!6}{GWDEG} & \cellcolor{gray!6}{1} & \cellcolor{gray!6}{NA} & \cellcolor{gray!6}{1} & \cellcolor{gray!6}{NA}\\
\midrule
\midrule
Outcome Homophily & 1 & NA & 1 & NA\\
\cellcolor{gray!6}{Intercept} & \cellcolor{gray!6}{1} & \cellcolor{gray!6}{1} & \cellcolor{gray!6}{1} & \cellcolor{gray!6}{1}\\
\addlinespace
Treatment & 1 & 1 & 1 & 1\\
\cellcolor{gray!6}{Neighbors Treated} & \cellcolor{gray!6}{0.1} & \cellcolor{gray!6}{0.1} & \cellcolor{gray!6}{0.1} & \cellcolor{gray!6}{0.1}\\
Neighbors Outcomes & 0.1 & 0.1 & 0.1 & 0.1\\
\midrule
\midrule
\cellcolor{gray!6}{Stochastic Edges} & \cellcolor{gray!6}{Yes} & \cellcolor{gray!6}{No} & \cellcolor{gray!6}{Yes} & \cellcolor{gray!6}{No}\\
Separable Likelihood & No & NA & Yes & NA\\
\bottomrule
\end{tabular}
\end{table}

The ERNM includes a mild spill over through a peer treatment effect, and
contagion through a peer outcome effect, where in addition to homophily,
peer outcomes and treatments also increase the chance of a positive
nodal outcome. We also include a homophilous geometrically weighted
edgewise shared partner (GWESP) term on outcome, that is a GWESP term
where edgewise shared partners are only credited if they match on
outcome. This term represents outcome transitivity. We suggest that in
many cases a researcher would often believe that these effects are
present in a social network formation process, and would fit such an
ERNM to observed data.

The MRF formulation assumes a fixed network with parameters for outcome
GWESP, outcome homophily, number of treated neighbours, positive outcome
neighbours, main effect and intercept. This may represent a model that a
researcher assuming a fixed network, with the simplistic chain graph
approximation may adopt. Note that the MRF model can include terms that
are functions of both edges and nodes, e.g., outcome GWESP and outcome
homophily, but the calculation of these statistics only changes due to
the nodes changing, not the edges changing.

The ERGM augmented with logistic regression accounts for network
uncertainty with the ERGM, and the nodal outcomes with the logistic
regression.

We also consider a pure logistic regression model where the network is
only allowed for through the neighbour covariates in the logistic
regression.

\subsection{Results}

The causal estimands we consider are the treatment main effect, \(1\) to
\(5\) peer outcome effects, \(1\) to \(5\) peer treatment effect. We
simulate \(100\) network realizations from the ERNM. For each of these
realizations we generated samples from the posterior parameter
distribution for each of the \(4\) models. For each of the \(400\)
posterior parameter distributions we sampled \(100\) parameter
realizations and estimated the causal estimands for those realizations.
Thus the output of the simulation was \(100\) simulated networks with
\(4\) posterior distributions for each network, for each of the causal
estimands.

We note that this was a computationally demanding simulation. For each
of \(4\) DGPs, we fit \(100\) posterior distributions to simulated
networks from the ERNM. For each of these \(100\) distributions we then
drew \(100\) samples from the posterior, for each of which we simulated
\(100\) networks to infer the missing potential outcomes. The fitting of
each of the posterior distributions typically required of the order of
\(10^4\) burn-in simulations with each step requiring a new ERNM MCMC,
which required a toggle burn-in of order \(10^4\). So each posterior
fitting procedure required, the \(10^8\) ERNM toggles with associated
change statistic calculation. As there were \(4\times 10^2\) posterior
distribution required to be fit, the posterior fitting step required
\(4\times 10^{10}\) ERNM network toggles. Simulating and inferring the
missing potential outcomes also requires MCMC burn-ins, though as
multiple steps were not required it is a lower order component of the
computation time.

Ordinarily the researcher would observe one network, fit one posterior
and simulate networks to infer the causal effect, which is feasible for
networks of the order of hundreds of nodes.

Table \ref{tab:ERNM_coverage} shows the mean posterior-mean and the
Frequentist coverage rates of the \(95\%\) Bayesian credible intervals,
together with the true causal estimands of the DGP. The coverage rates
are included to enable calibration of the credible intervals
\citep{Little2011}. We also show the mean mean-a-posteriori to justify
that, on average, the posteriors are centred around the true value.

We note that the ERGM logistic and pure logistic models recover some
outcome effects on average, but perform very poorly on treatment
effects. The ERNM and MRF posteriors seem to broadly be centred close to
the true effects, though the MRF posteriors have much lower Frequentist
coverage than the ERNM model, perhaps suggesting optimistically low
variance in the posterior distribution. This is as expected as the MRF
model does not account for randomness in the edges of the network. In
addition the MRF model was unable to identify higher order peer
treatment effects, denoted as NA in table \ref{tab:ERNM_coverage}. This
is because \(4\) and \(5\) peer treatments were not observed in any of
the simulated networks.

\begin{table}

\caption{\label{tab:unnamed-chunk-2}\label{tab:ERNM_coverage} Mean mean a-posteriori causal estimands fitted to 100 network simulations from ERNM. The coverage of the true mean by the 100 estimated 95\% credible intervals is shown in brackets.}
\centering
\begin{tabular}[t]{lrllll}
\toprule
  & True & ERNM & MRF & ERGM+Logistic & Logistic\\
\midrule
\cellcolor{gray!6}{main} & \cellcolor{gray!6}{0.28} & \cellcolor{gray!6}{0.27 (65\%)} & \cellcolor{gray!6}{0.28 (67\%)} & \cellcolor{gray!6}{-0.03 (0\%)} & \cellcolor{gray!6}{-0.17 (0\%)}\\
1-peer-out & 0.28 & 0.27 (69\%) & 0.36 (31\%) & 0.18 (59\%) & 0.16 (8\%)\\
\cellcolor{gray!6}{2-peer-out} & \cellcolor{gray!6}{0.50} & \cellcolor{gray!6}{0.5 (68\%)} & \cellcolor{gray!6}{0.64 (21\%)} & \cellcolor{gray!6}{0.45 (98\%)} & \cellcolor{gray!6}{0.39 (53\%)}\\
3-peer-out & 0.66 & 0.65 (63\%) & 0.77 (34\%) & 0.66 (97\%) & 0.58 (72\%)\\
\cellcolor{gray!6}{4-peer-out} & \cellcolor{gray!6}{0.77} & \cellcolor{gray!6}{0.74 (57\%)} & \cellcolor{gray!6}{0.82 (52\%)} & \cellcolor{gray!6}{0.76 (95\%)} & \cellcolor{gray!6}{0.7 (77\%)}\\
\addlinespace
5-peer-out & 0.82 & 0.8 (58\%) & 0.83 (62\%) & 0.8 (95\%) & 0.76 (81\%)\\
\cellcolor{gray!6}{1-peer-treat} & \cellcolor{gray!6}{0.13} & \cellcolor{gray!6}{0.14 (80\%)} & \cellcolor{gray!6}{0.16 (69\%)} & \cellcolor{gray!6}{-0.06 (0\%)} & \cellcolor{gray!6}{0 (11\%)}\\
2-peer-treat & 0.27 & 0.27 (77\%) & 0.28 (70\%) & -0.12 (0\%) & 0 (13\%)\\
\cellcolor{gray!6}{3-peer-treat} & \cellcolor{gray!6}{0.39} & \cellcolor{gray!6}{0.39 (71\%)} & \cellcolor{gray!6}{0.4 (70\%)} & \cellcolor{gray!6}{-0.16 (0\%)} & \cellcolor{gray!6}{0 (14\%)}\\
4-peer-treat & 0.50 & 0.48 (72\%) & NA & -0.2 (0\%) & 0.01 (15\%)\\
\addlinespace
\cellcolor{gray!6}{5-peer-treat} & \cellcolor{gray!6}{0.56} & \cellcolor{gray!6}{0.54 (70\%)} & \cellcolor{gray!6}{NA} & \cellcolor{gray!6}{-0.23 (1\%)} & \cellcolor{gray!6}{0.01 (14\%)}\\
\bottomrule
\end{tabular}
\end{table}

However if we work consistently in the Bayesian framework, for the
networks were generated from an ERNM, the posterior causal estimand
distribution derived from the ERNM posterior, is the ``ground truth'' in
the Bayesian sense. Thus the correct assessment of the performance of
any given method should be comparing its posterior causal estimand
distribution to the ground truth distribution. The comparison for a
given method is to compare its posterior fit based on each of the
\(100\) simulated networks to the corresponding posterior derived from
the true DGP. Therefore understanding the performance of each method
reduces to comparing distributions. We use the relative distribution
\citep{handcock1999} to this end. We consider the relative rank
distribution of each of the pairs of models, using boundary adjusted
kernel density estimation using the \texttt{reldist} package
\citep{reldist_package}.

Table \ref{tab:ERNM_KL} shows the estimated Kullback-Leibler (KL)
divergences between the relative rank distribution and the uniform
distribution. To calibrate the size of the divergences, the KL
divergence between two unit variance Gaussian distributions is equal to
one-half the squared difference between their means. On this scale, a KL
divergence of \(d\) corresponds to a \(\sqrt{2d}\) mean difference. As
the ERNM model is being compared against itself, the expect the
divergence to be \(0\). The others have large KL divergences from the
ERNM posterior, suggesting that they are not able to recreate the true
posterior distribution of important causal estimands when misspecified.

\begin{table}

\caption{\label{tab:unnamed-chunk-4} \label{tab:ERNM_KL} Mean KL divergence of relative rank distributions of posteriors for causal estimands across 100 network simulations from the ERNM}
\centering
\begin{tabular}[t]{lrlrr}
\toprule
  & ERNM & MRF & ERGM+Logistic & Logistic\\
\midrule
\cellcolor{gray!6}{main} & \cellcolor{gray!6}{0} & \cellcolor{gray!6}{1.09} & \cellcolor{gray!6}{3.77} & \cellcolor{gray!6}{4.44}\\
1-peer-out & 0 & 2.45 & 2.18 & 3.33\\
\cellcolor{gray!6}{2-peer-out} & \cellcolor{gray!6}{0} & \cellcolor{gray!6}{2.59} & \cellcolor{gray!6}{1.18} & \cellcolor{gray!6}{1.88}\\
3-peer-out & 0 & 2.2 & 1.08 & 1.26\\
\cellcolor{gray!6}{4-peer-out} & \cellcolor{gray!6}{0} & \cellcolor{gray!6}{1.56} & \cellcolor{gray!6}{1.03} & \cellcolor{gray!6}{1.13}\\
\addlinespace
5-peer-out & 0 & 1.22 & 0.93 & 1.08\\
\cellcolor{gray!6}{1-peer-treat} & \cellcolor{gray!6}{0} & \cellcolor{gray!6}{1.24} & \cellcolor{gray!6}{4.23} & \cellcolor{gray!6}{3.52}\\
2-peer-treat & 0 & 1.25 & 4.24 & 3.52\\
\cellcolor{gray!6}{3-peer-treat} & \cellcolor{gray!6}{0} & \cellcolor{gray!6}{1.34} & \cellcolor{gray!6}{4.23} & \cellcolor{gray!6}{3.48}\\
4-peer-treat & 0 & NA & 4.21 & 3.44\\
\addlinespace
\cellcolor{gray!6}{5-peer-treat} & \cellcolor{gray!6}{0} & \cellcolor{gray!6}{NA} & \cellcolor{gray!6}{4.19} & \cellcolor{gray!6}{3.41}\\
\bottomrule
\end{tabular}
\end{table}

\section{Case-Study of Smoking Behavior within a High School}\label{sec:example}

In this section we give a real data case-study utilizing ERNM to relax
the fixed network assumption as well as conditional unconfoundedness of
the edges and nodal potential outcomes. In this case we do not know the
ground truth, so the purpose of this section is to demonstrate that our
method produces plausible posterior estimand distributions and to
highlight the differences between methods for these data. We also
performed a simulation study with known data generating processes which
is contained in the supplement.

We consider one of the school social networks from the National
Longitudinal Study of Adolescent Health \citep{AddHealth2007}. The
network we used for this example has \(869\) nodes of which \(462\) were
male and \(407\) were female, with \(344\) having reported trying a
cigarette at least once. For consistency with \cite{fellows2012}, gender
was coded as \(1\) for male and \(0\) for females, hence the gender
coefficients reported correspond to males.

We consider the following estimands and estimate them under different
frameworks:

\begin{enumerate}
\item $k$-peer effect of the gender of peers on smoker status of the ego 
\item $k$-peer outcome effect of peer smoker status on smoker status of the ego.
\end{enumerate}

Within our Bayesian framework we consider the following models for
imputing the required potential outcomes to claim causal inference:

\begin{enumerate}
\item Full ERNM model with potential outcome as a random nodal covariates.
\item Markov random field model with fixed network
\item Logistic regression.
\end{enumerate}

In our framing the outputs are posterior distributions of causal
estimands. For information we also show the results of the MRF model,
with parameters estimated through maximum pseudo-likelihood estimation
and potential outcomes derived from these as in \cite{tchetgen2020}. In
line with the known bias of pseudo likelihood estimates for ERGM
\citep{vanDuijn2009} we believe this method will perform poorly.

We used a version of the exchange algorithm \citep{Murray2006}, with an
extension which allows for efficient sampling from the ERNM posterior.
The development is given in the supplement. While informative priors are
compatible with the computational framework, here we report based on a
uniform prior over all parameters. The results do not appear to be
sensitive to the choice of prior.

\begin{table}

\caption{\label{tab:unnamed-chunk-6}\label{tab:add_health_models}
      Summary of posterior distributions of network models, posterior means are shown with posterior standard errors in parentheses}
\centering
\begin{tabular}[t]{llll}
\toprule
  & ERNM & MRF & Logistic Regression\\
\midrule
\cellcolor{gray!6}{Edges} & \cellcolor{gray!6}{-4.93 (0.03)} & \cellcolor{gray!6}{NA} & \cellcolor{gray!6}{NA}\\
Grade GWESP & 0.11 (0) & NA & NA\\
\cellcolor{gray!6}{GWDEG} & \cellcolor{gray!6}{-1.67 (0.36)} & \cellcolor{gray!6}{NA} & \cellcolor{gray!6}{NA}\\
Grade Homophily & 4.13 (0.05) & NA & NA\\
\cellcolor{gray!6}{Sex Homophily} & \cellcolor{gray!6}{0.67 (0.08)} & \cellcolor{gray!6}{NA} & \cellcolor{gray!6}{NA}\\
\addlinespace
Smoke Homophily & 0.48 (0.06) & 0.43 (0.07) & NA\\
\midrule
\midrule
\cellcolor{gray!6}{Intercept} & \cellcolor{gray!6}{0.95 (0.11)} & \cellcolor{gray!6}{0.75 (0.12)} & \cellcolor{gray!6}{-1.07 (0.2)}\\
Gender & -0.07 (0.03) & 0.02 (0.04) & 0.47 (0.16)\\
\cellcolor{gray!6}{Female neighbors} & \cellcolor{gray!6}{0.11 (0.02)} & \cellcolor{gray!6}{0.08 (0.02)} & \cellcolor{gray!6}{-0.17 (0.05)}\\
Smoker neighbours & -0.64 (0.12) & -0.46 (0.12) & 0.54 (0.06)\\
\midrule
\midrule
\addlinespace
\cellcolor{gray!6}{Stochastic Edges} & \cellcolor{gray!6}{Yes} & \cellcolor{gray!6}{No} & \cellcolor{gray!6}{No}\\
Stochastic Covariates & Smoker Status & Smoker Status & Smoker Status\\
\cellcolor{gray!6}{Separable Likelihood} & \cellcolor{gray!6}{No} & \cellcolor{gray!6}{NA} & \cellcolor{gray!6}{NA}\\
\bottomrule
\end{tabular}
\end{table}

Table \ref{tab:add_health_models} gives a summary of the posterior
distributions for each of the models, showing the posterior means with
the posterior standard errors in parentheses. We note that parameters
should not be compared across models, as the functional forms are
different, we show this table to summarize the posteriors, but to also
highlight the differences between the models.

For GWESP and geometrically weighted degree (GWDEG) terms decay
parameters were fixed at \(0.5\). The use of these geometrically
weighted terms is in part necessary to avoid degeneracy issues
\citep{handcock2003, snijders2006}, but also implicitly induces the
social circuit dependence assumption for the ERGM, rather than the more
restrictive Markov assumption. However we used the ERNM style homophily
terms \citep{fellows2012} for both the ERNM and ERGM model, which in
fact induce non local dependencies. Thus the dependence structures of
the edges in these models are unknown and best described as ``complex''.
In addition we enforce homogeneity in school grade for the edgewise
shared partners in the GWESP term as there is a very strong grade
structure to transitivity in this network.

We do not interpret the posterior parameter distributions directly,
rather we make comparison through the smoker peer effect on the smoker
status of a node.

\begin{figure}

{\centering \includegraphics[width=1\linewidth]{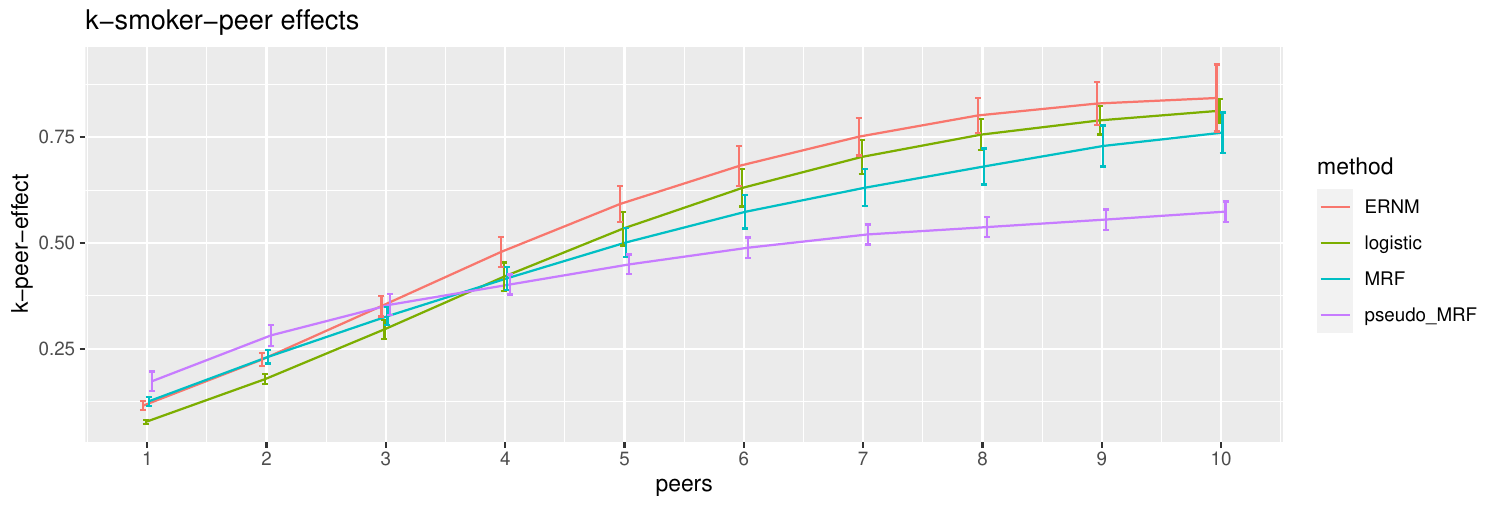} 

}

\caption{\label{fig:k_peer} Plot comparing the posteriors distribution of the ERNM, MRF and logistic regression estimated k-peer outcome effect. The MRF with parameters equal to the maximum pseduo likelihood esimtate is also shown}\label{fig:unnamed-chunk-10}
\end{figure}

Table \ref{tab:k_peer} and Figure \ref{fig:k_peer} show the the
\(k\)-smoker-peer effect estimates. These are is estimated as the
additional chance of smoking that having \(k\) smoker friends has over
having no smoker friends. The ERNM and MRF model are in agreement for
peers one to three, with some divergence after this. The logistic
regression model is markedly different from the ERNM for one and two
peer effects, while for higher effects the estimates are closer to the
ERNM estimates. The pseudo likelihood estimated MRF model, as expected,
are quite different. This helps confirms our prior belief that this
estimation method is likely biased in the Frequentist sense,
consequently the fitted model does not fit the data well, and does
accurately estimate causal estimands.

We believe the ERNM to be most plausible from a theoretical perspective.
Whilst in this example the effect size difference from the MRF model was
not large, we believe it to be a more robust approach when estimating
network causal effects. In particular where there is strong transitivity
interacting with nodal outcomes as well as for smaller networks, we
expect the effect would be larger.

\begin{figure}

{\centering \includegraphics[width=1\linewidth]{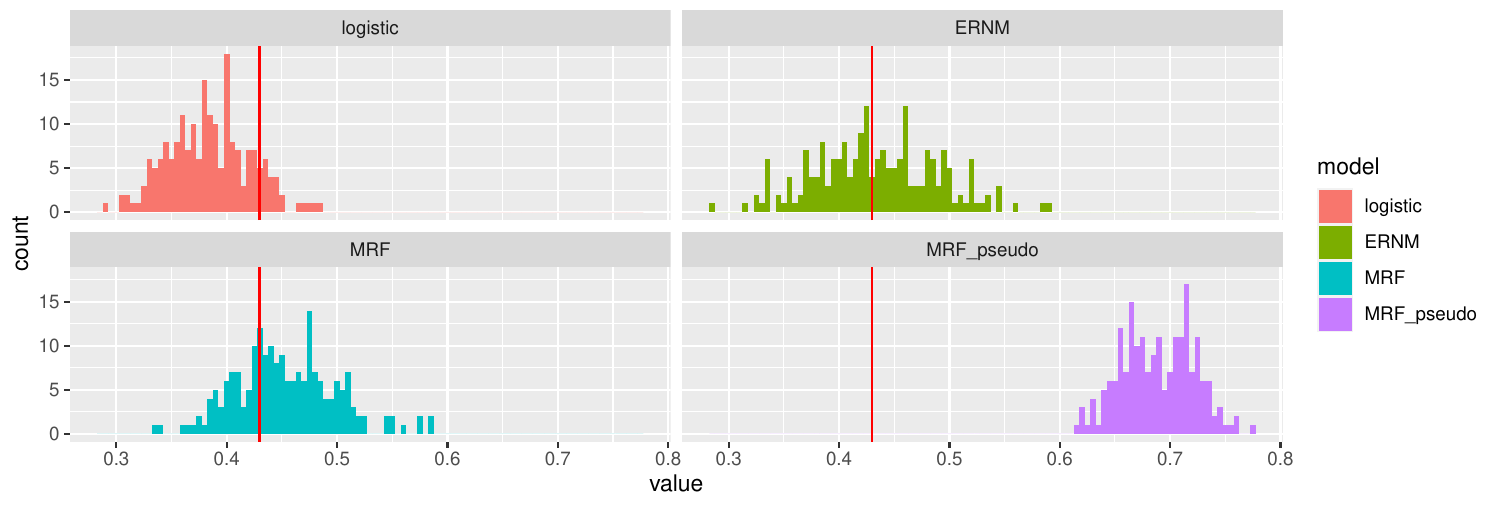} 

}

\caption{\label{fig:triad_smoke} Distribution of proportion of triads with all nodes smokers. ERNM and MRF posterior simulated distributions fit the observed data considerably better that logistic regression or the pseudo-likelihood estimated MRF.}\label{fig:unnamed-chunk-12}
\end{figure}

\begin{figure}

{\centering \includegraphics[width=1\linewidth]{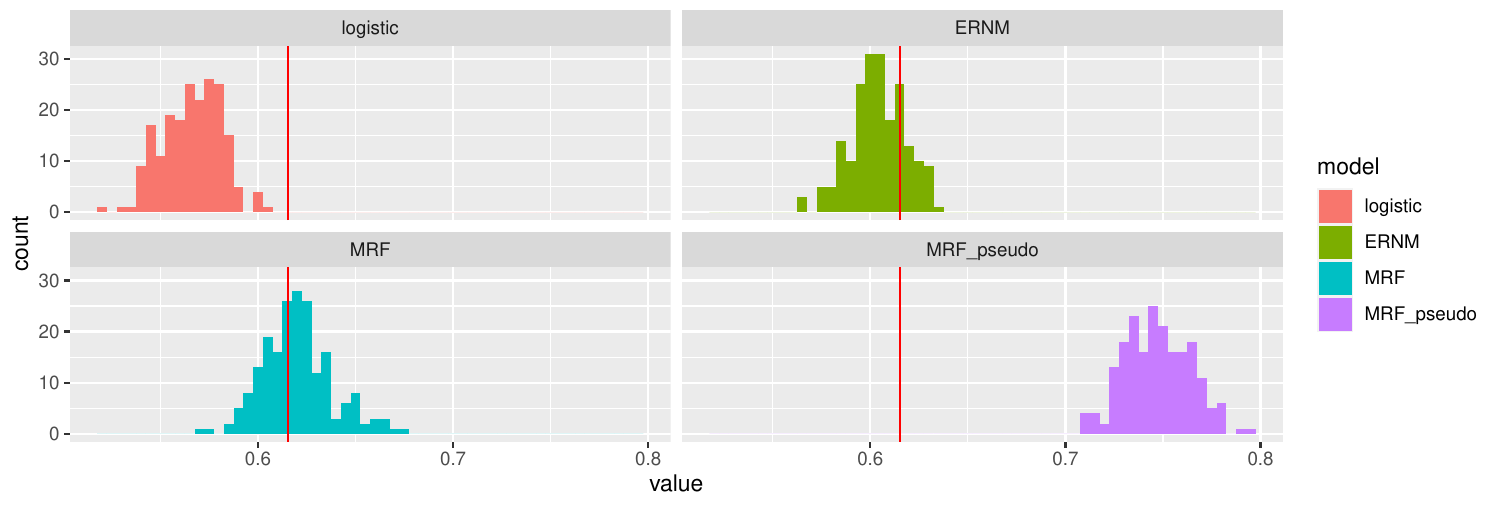} 

}

\caption{\label{fig:edge_smoke} Distribution of proportion edges with both nodes smokers. ERNM and MRF posterior simulated distributions fit the observed data considerably better that logistic regression or the pseudo-likelihood estimated MRF.}\label{fig:unnamed-chunk-13}
\end{figure}

In the context of our problem, the particular advantage of ERNM is that
for simulated networks the smoker nodes are observed within network
sub-structures consistent with the observed network. Figures
\ref{fig:edge_smoke} and \ref{fig:triad_smoke} compare the distributions
of the proportion of smoker edges and triads in the networks simulated
from each model. We consider the proportion of smoker triads, as all
models' simulations underestimate the absolute number of triads. We note
that the ERNM model and the full MRF fits considerably better that the
pseudo MRF and the logistic regression model.

We show Bayesian posterior prediction goodness-of-fit graphics in the
style of \cite{Hunter_Goodreau_2008} in the supplement. These
demonstrate that networks simulated from the ERNM model posterior,
correspond closely to the observed data.

\begin{table}

\caption{\label{tab:unnamed-chunk-8}\label{tab:k_peer} Posterior means of the k peer smoker outcome effect ATEs, for various methods. The pseudo_MRF value is the
mean simulated from the parameter estimate.}
\centering
\begin{tabular}[t]{llll}
\toprule
ERNM & MRF & pseudo\_MRF & Logistic Regression\\
\midrule
\cellcolor{gray!6}{0.12 (0.01)} & \cellcolor{gray!6}{0.13 (0.01)} & \cellcolor{gray!6}{0.17 (0.02)} & \cellcolor{gray!6}{0.08 (0)}\\
0.22 (0.02) & 0.23 (0.02) & 0.28 (0.02) & 0.18 (0.01)\\
\cellcolor{gray!6}{0.35 (0.02)} & \cellcolor{gray!6}{0.33 (0.02)} & \cellcolor{gray!6}{0.35 (0.02)} & \cellcolor{gray!6}{0.3 (0.02)}\\
0.48 (0.04) & 0.42 (0.03) & 0.4 (0.02) & 0.42 (0.03)\\
\cellcolor{gray!6}{0.59 (0.04)} & \cellcolor{gray!6}{0.5 (0.03)} & \cellcolor{gray!6}{0.45 (0.02)} & \cellcolor{gray!6}{0.53 (0.04)}\\
\addlinespace
0.68 (0.05) & 0.57 (0.04) & 0.49 (0.02) & 0.63 (0.04)\\
\cellcolor{gray!6}{0.75 (0.04)} & \cellcolor{gray!6}{0.63 (0.04)} & \cellcolor{gray!6}{0.52 (0.02)} & \cellcolor{gray!6}{0.7 (0.04)}\\
\bottomrule
\end{tabular}
\end{table}

\setkeys{Gin}{width=\textwidth}

\section{Discussion}\label{sec:discussion}

In this paper we model causality when the underlying population is
networked and that network endogenous to the social process. Our
approach jointly stochastically models the links and nodal covariates in
the network, better representing our state of knowledge and their
codependency.

Considering a DAG, we suggest that the estimating true causal effects in
this setting, is almost always intractable due to the usual lack of fine
grained temporal data, as well as the highly complex causal structure.
We present a chain graph approximation to the DAG, which allows for a
dependence structure that we believe to be compatible with such
problems. We then frame the approximate causal inference in terms of
network equilibrium potential outcomes, that is, potential outcomes that
are free to depend on nodes in the neighbourhood of the node in
question. We propose the use of ERNMs to jointly model both random edges
and nodal covariates. We also develop a simple modification to the
exchange algorithm allowing for a feasible sampling from the posterior
distribution. We use the posterior distribution, through simulation, to
impute the distribution of the missing potential outcomes, allowing the
consideration of the distribution of the causal estimands. We showed,
using a school network from the National Longitudinal Study of
Adolescent Health, that failing to account for the network structure of
the problem could lead to misleading qualitative conclusions, in
particular when considering the one- and two- peer outcome effect.

Our primary contributions lie in considering the consequences of
relaxing of the commonly made fixed network assumption and proposing the
use of suitable social network models to estimate causal effects. This
relaxation complicates the causal structure considerably and
necessitates the use of complex models to derive causal estimands.
Clearly the relaxation of this assumption also allows for greater
generalizability. Our inferences hold for the given node set and social
process, whereas assuming the fixed network narrows the scope to that
observed network only. Our method is only applicable to the given node
set. However we suggest that the posteriors derived from the given
network can serve as strong priors for ``similar'' networks. In general,
qualitative features of the posteriors from the given network can useful
inform other analyses. It is not possible to make further statements
than this for networks of different sizes.

We believe that in the context of social network analysis, where
individual attributes are heavily influenced by social context, such
covariation of edges and nodal covariates is overwhelmingly more
representative of many social processes. That is, in many social
networks we believe it highly likely that individual characteristics and
the connections that form between individuals are strongly dependent.
This is especially true if the network evolves over time. We also note
that avoiding such issues by letting the size of the network become
large and invoking asymptotic arguments, fundamentally mis-interprets
the problem. The phenomenon of interest; the inter dependencies of
actors, is a result of the small size of the network, for example
dependence assumptions implicitly made in modelling a \(30\) node
network usually do not make sense to apply to say a network of \(3000\)
nodes. Thus arguments that rely on the number of nodes being large, are
incoherent, as they rely on changing the structure of the problem
itself, to understand uncertainty. As our approach is fully Bayesian on
the fixed node set, we do not require asymptotic arguments on the number
of nodes in a network.

Notable by its absence is discussion of suitable prior distributions for
ERNMs. We note there has been some work developing conjugate prior
distributions for ERGMs \citep{Ranran2011} and strongly suspect that
such an approach may also be applicable in our setting. In practice flat
Gaussian prior are often used for ERGMs \citep{caimo2011}. For the
purpose of demonstrating our approach, we used uniform priors, which
make no account for the geometry of exponential families, but allow us
defer careful consideration of possible priors to future work, while
demonstrating the utility of our method. We note that we performed
similar posterior fits with flat Gaussian priors, which did not effect
the posteriors substantially.

The cost of our approach is the strong assumption that such a complex
process, can be adequately modelled by an ERNM. This is in general the
main criticism of model based causal inference approaches, that models
are mis-specified with unknown consequences. In a network setting this
mis-specification is often acute e.g.~constant marginal effect of
additional smoker friends in the linear potential outcomes model. Our
central argument is that a complex model is much less mis-specified than
current approaches. We have sought to justify this with real data and
simulations, though propose this as a future area of research. For
example, how dependent do outcomes in networks need to be to invalidate
conclusions made with unrealistic models?

The mis-specification may seem to be cause for pessimism, however we
emphasize that network settings are indeed the extreme case of small
data, as we usually only have one observations of a network on a fixed
set of nodes. Thus intuitively we should expect strict functional form
assumptions to be required to generate any meaningful statistical, and
especially causal inference. In fact we argue that approaching network
problems with simpler assumptions is problematic, whilst potentially
less prone to mis-specification in the sense that simple models can be
used, this easily glosses over the inherent difficulty of dealing with
network data where nodes and edges are strongly dependent on other nodes
and edges.

we also note that specifying a model for the full network data
generation process also allows inference in cases where only a subset of
the network is sampled. Accounting for such sampling structure is likely
analogous to the method for ERGMs in \cite{gile2016}. Accounting for
this is not possible with the other methods considered in this work. In
addition the network process can be considered to evolve after treatment
conditional on some pre treatment network. Such an approach may lead to
increased power, for randomised control trials on a networked
population, at the cost of our modelling assumptions.

We believe that meaningful steps can be made towards causal inference on
networks, through careful consideration of the complex causal structure
of such problems. Whilst we make strict assumption on the function form
of this, if the researcher is unwilling to make such assumptions, we
opine casual inference is out of reach. We suggest it is better to
acknowledge the complexity of the situation, and therefore claim that
causal inference is not possible, than employing highly restrictive
assumptions on the dependence structure of the data generating process,
to allow simpler models to be employed.

\section{Acknowledgements}

This article is based upon work supported by the National Science
Foundation(NSF, MMS-0851555, SES-1357619, IIS-1546259) and National
Institute of Child Health and Human Development (NICHD, R21HD063000,
R21HD075714 and R24-HD041022). The content is solely the responsibility
of the authors and do not necessarily represent the official views of
the National Institutes of Health or the National Science Foundation.

\section{Supplementary Materials}

\begin{description}

\item[Supplement] The supplement contains an additional chain graph approximation diagram, a review of ERNM and Bayesian computation for them as well as an MCMC convergence analysis for the adolescent health network. (pdf) 

\end{description}

\bibliographystyle{chicago}
\bibliography{bib_causal}

\end{document}


\newtheorem{theorem}{Theorem}
\newtheorem{example}{Example}
\newtheorem{lemma}{Lemma}
\newtheorem{proposition}{Proposition}
\newtheorem{definition}{Definition}
\newtheorem{assumption}{Assumption}
\newcommand{\R}{\mathbb{R}}
\newcommand{\N}{\mathbb{N}}
\newcommand{\E}{\mathbb{E}}
\newcommand{\V}{\mathbb{V}}
\newcommand{\bfR}{\mathbf{R}}
\newcommand{\bfX}{\mathbf{X}}
\newcommand{\bfW}{\mathbf{W}}
\newcommand{\bfD}{\mathbf{D}}
\newcommand{\INT}{\int_{-\infty}^{+\infty}}
\newcommand{\p}{\partial}
\newcommand{\ra}{\Rightarrow}
\newcommand{\dH}{d\mathscr{H}}
\newcommand{\ch}{\text{cosh}}
\newcommand{\sh}{\text{sinh}}
\newcommand{\ex}{\mathbb{E}\left[X\right]}
\newcommand{\ey}{\mathbb{E}\left[Y\right]}
\newcommand{\indep}{\perp \!\!\! \perp }
\newcommand{\blandscape}{\begin{landscape}}
\newcommand{\elandscape}{\end{landscape}}

\setcounter{secnumdepth}{4}

\section{Chain Graph Introduction}

This section gives a very brief introduction to chain graphs, to help
understand why they are appropriate for approximating the structure of a
social process that evolves over time. This section exclusively repeats
material from \cite{frydenberg1990}, \cite{lauritzen2002}, and is
included for completeness as the entire article relies on chain graph
concepts.

\subsection{Graph Definitions}

We first define a graph \(\mathscr{G}\) as a set
\(\lbrace V,E(V) \rbrace\) where \(V\) is a finite set of vertices, and
\(E(V)\) is a set of edges between the vertices contained in \(V\), both
directed \(\rightarrow\) and undirected \(-\) edges are permitted. We
denote the subgraph induced by a vertex \(a\) as \(\mathscr{G}_{a}\)

A partially directed cycle is defined as subset of
\(\lbrace v_{i},\ldots ,v_{k}\rbrace \subseteq V\), re-indexing as
necessary, such that for each \(i\), \(v_{i}\rightarrow v_{i+1}\) or
\(v_{i}-v_{i+1}\) with at least one of the edges directed. A chain graph
is defined as a graph with at no partially directed cycles.

There exists a path between \(u\) and \(v\) if \(u=v\) or if \(u\) and
\(v\) are connected by a sequence of edges through vertices
\(\lbrace v_1,\ldots ,v_k \rbrace\). A path is directed if any of edges
between \(v_i\) and \(v_{i+1}\) are directed and in the same direction
i.e.~\(v_{i}\) to \(v_{i+1}\) or \(v_{i}\) to \(v_{i-1}\).

For vertices \(v,u \in V\) we define \(u\) as a parent of \(v\) if
\(u \rightarrow v\) and write \(u \in pa(v)\) for undirected edges we
define \(u\) and \(v\) neighbours if \(u-v\) and write \(u \in ne(v)\)
and \(v\in ne(u)\). The boundary of \(v\) is defined as the union of the
neighbours and the parents, the closure \(cl(v)\) is the
\(v \cup bd(v)\)

A subgraph induced by a vertex set \(A\) is denoted \(\mathscr{G}_{A}\)
and contains all vertices in \(A\) and all edges involving vertices in
\(A\). For vertices \(a\) and \(b\) minimal complex is an induced
subgraph of the form \(a\rightarrow v_1 - \dots - v{r} \leftarrow b\).

The chain components \(\mathscr{T}\) of a chain graph \(G\) are the
connected components of \(G\) with all directed edges deleted.

We define the future \(\phi(U)\) and past \(\pi(U)\) for a vertex \(u\)
as
\(\phi(u) = \lbrace v\in V \vert \exists \text{ a path from } u \text{ to } v \rbrace\)
and
\(\pi(u) = \lbrace v\in V \vert \exists \text{ a path from } v \text{ to } u \rbrace\).
The future and past of a vertex set \(U\) is defined as the union of the
futures and past of all the nodes in the set. A set is terminal if
\(\phi(U) = \emptyset\) and initial if \(\pi(U) = \emptyset\).

A subset of vertices \(U\) is anterior if it can be generated by
removing terminal chain components. This allows us to finally define the
moral graph, which will be used to define the Markov property for chain
graphs. The moral graph \(\mathscr{G}^{m}\) is the graph formed by
considering the chain components of the graph, and completing their
boundaries, i.e.~replacing directed edges with undirected edges between
chain components and their parents, and connecting all parents with
undirected edges.

\subsection{Chain Graph Markov Property}

Graphical models serve to encode the structure of conditional
independence relationships between variables. The Markov property of a
graphical model is this encoding. The chain graph Markov property was
defined and investigated in \cite{frydenberg1990}.

We restate Theorem 3.3 of \cite{frydenberg1990} which establishes the
equivalence of 4 proposed Markov properties.

\begin{theorem}{Chain Graph Markov Property Equivalences Theorem 3.3 of \cite{frydenberg1990}}

For a chain graph $\mathscr{G}$ and probability measure $P$ such that:

\begin{align*}
(A \perp B \vert D\cup C \quad \text{and} \quad A\perp C \vert D \cup B) \quad \text{implies} \quad A\perp B \cup C \vert D 
\end{align*}

The following are equivalent:

\begin{enumerate}
\item Global $\mathscr{G}$-Markovian : $A \perp B \vert C$ whenever $C$ separates $A$ and $B$ in $\left( \mathscr{G}_{an(A \cup B \cup C)} \right)$ - the moralization of the anterior set of the union.

\item Local $\mathscr{G}$-Markovian : $u \perp [V \setminus\phi(u)]\setminus cl(u) \vert bd(u) \quad \forall \quad u$  
\item Pairwise $\mathscr{G}$-Markovian  $u \perp v \vert  [V \smallsetminus\phi(u)]\smallsetminus \lbrace u,v \rbrace$ for $v\not\in \phi(u)$, and $u$ and $v$ not adjacent

\item $\mathscr{G}_{A}^{m}$-Markovian for every anterior set $A$ of $\mathscr{G}$. Where $\mathscr{G}_{A}^{m}$-Markvoian refers to the undirected Markov property in \cite{frydenberg1990}
\end{enumerate}
\end{theorem}

These Markov properties lead to the factorization of all chain graph
models in Theorem 4.1 of \cite{frydenberg1990}.

\begin{theorem}{Chain Graph Factorizations (Partial Version of Theorem 4.1 of \cite{frydenberg1990})}

For a distribution $P$ with positive density $p$ w.r.t. some product measure $\mu = \bigtimes_{v\in V} \mu_{v}$ on space $\mathscr{H}$. If $P$ is $\mathscr{G}$-Markovian then p can be factorized as :

\begin{align}
p(x) = \prod_{\tau\in\mathscr{T}} \prod_{C\in\mathscr{C}_{\tau}} \psi_{\tau}^{C}\left(X_{C}\right)
\end{align}

Where $\mathscr{T}$ denotes the set of chain components in $\mathscr{G}$ and $\mathscr{C_{\tau}}$ denotes the set of cliques in $\left( \mathscr{G}_{cl(\tau)} \right) ^{m}$
\end{theorem}

This theorem provides that factorisation of chain graphs as described in
\cite{lauritzen2002} as a ``DAG of boxes'' where the boxes are the set
of chain components.

\subsection{Chain Graph Example Comparison with DAG}

In this section we follow the example given in Section \(5\) of
\cite{lauritzen2002}. We give a simple example of conditional dependence
structures that cannot be represented with a DAG, but can be with a
chain graph.

\begin{figure}
\begin{center}
\begin{subfigure}[t]{0.5\textwidth}
   \centering
   \begin{tikzpicture}
    \node[text centered] (a) {$A$};
    \node[right = 1.5 of a, text centered] (c) {$C$};
    \node[below = 1.5 of a, text centered] (b) {$B$};
    \node[below = 1.5 of c, text centered] (d) {$D$};
    
    \draw[->, line width= 1] (a) -- (c);
    \draw [->, line width= 1] (b) -- (d);
    \draw[-,line width= 1] (c) --(d);
  \end{tikzpicture}
   \caption{Full chain graph}
   \label{fig:chain_1}
\end{subfigure}\hfill
\begin{subfigure}[t]{0.5\textwidth}
   \centering
    \begin{tikzpicture}
    \node[text centered] (a) {$A$};
    \node[right = 1.5 of a, text centered] (c) {$C$};
    \node[below = 1.5 of a, text centered] (b) {$B$};
    \node[below = 1.5 of c, text centered] (d) {$D$};
    
    \draw[-, line width= 1] (a) -- (c);
    \draw [-, line width= 1] (b) -- (d);
    \draw[-,line width= 1] (c) --(d);
    \draw[-,line width= 1] (a) --(b);
    \end{tikzpicture}
   \caption{Moralization of smallest anterior set containing $\lbrace a,b,c,d \rbrace$. Moralization, connects the largest chain component to its parents with undirected edges, and connects these parents with undirected edges also}
   \label{fig:chain_2}
\end{subfigure}
   \caption{Full chain graph example and moral graph}
\end{center}
\label{fig:chain}
\end{figure}
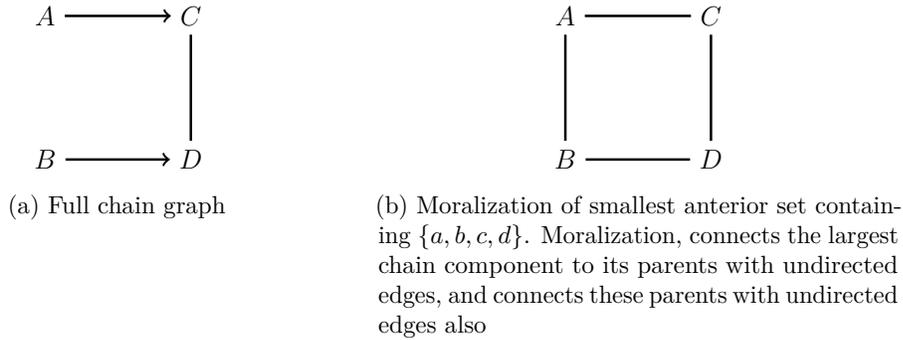

Figure \ref{fig:chain_1} shows a simple chain graph. The density can be
factorised under the chain graph factorisation theorem as
\(f(a,b,c,d) = f(c,d \vert a,b)f(a)f(b)\). However the chain graph
markov property places restrictions on the form of \(f(c,d\vert a,b)\).

Figure \ref{fig:chain_2} shows the transformation to the moralization of
the anterior set of the union, so we can use the global Markov property.
As the sets \(\lbrace b,c \rbrace\) and \(\lbrace a,d \rbrace\) separate
\(a\) and \(d\) and \(b\) and \(c\) respectively we have:

\begin{align}
a\perp d \vert \lbrace b,c \rbrace \qquad  b\perp c \vert \lbrace a,d \rbrace 
\end{align}

In addition using the pairwise Markov property and noting that
\(\phi(a) = \lbrace c,d \rbrace\) we have that:

\begin{align}
& a \perp b \vert  [V \setminus\phi(a)]\setminus \lbrace a,b \rbrace\\
\rightarrow & a \perp b \vert \emptyset \\
\rightarrow & a\perp b
\end{align}

\begin{figure}
\begin{center}
\begin{subfigure}[t]{0.5\textwidth}
   \centering
   \begin{tikzpicture}
    \node[text centered] (a) {$A$};
    \node[right = 1.5 of a, text centered] (c) {$C$};
    \node[below = 1.5 of a, text centered] (b) {$B$};
    \node[below = 1.5 of c, text centered] (d) {$D$};
    \node[below right = 0.75 of c, text centered] (p) {$Parent$};
    
    \draw[->, line width= 1] (a) -- (c);
    \draw [->, line width= 1] (b) -- (d);
    \draw [->, line width= 1] (p) -- (c);
    \draw [->, line width= 1] (p) -- (d);
  \end{tikzpicture}
   \caption{DAG with common parent}
   \label{fig:DAG_1a}
\end{subfigure}\hfill
\begin{subfigure}[t]{0.5\textwidth}
   \centering
    \begin{tikzpicture}
    \node[text centered] (a) {$A$};
    \node[right = 1.5 of a, text centered] (c) {$C$};
    \node[below = 1.5 of a, text centered] (b) {$B$};
    \node[below = 1.5 of c, text centered] (d) {$D$};
    \node[rectangle,draw](ch)[below right = 0.75 of c, text centered]  {$Child$};
    
    \draw[->, line width= 1] (a) -- (c);
    \draw [->, line width= 1] (b) -- (d);
    \draw [->, line width= 1] (c) -- (ch);
    \draw [->, line width= 1] (d) -- (ch);
    \end{tikzpicture}
\caption{DAG with conditioning on common child}
   \label{fig:DAG_1b}
\end{subfigure}
   \caption{Example DAGs representing different Markov properties to the Chain graph in Figure \ref{fig:chain_1}}
\end{center}
\end{figure}
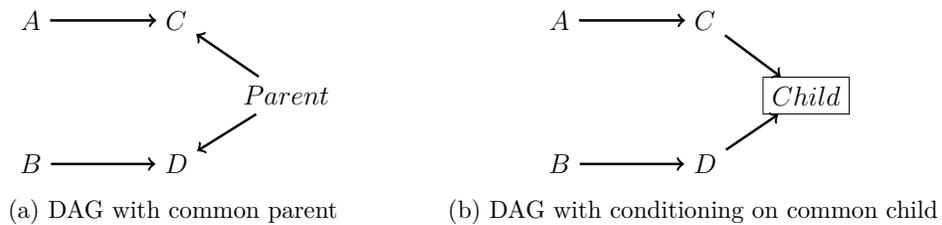

Figure \ref{fig:DAG_1} shows 2 DAGs that one might think encode the
Markov property as Figure \ref{fig:chain_1}. Figure \ref{fig:DAG_1a}
shows an additional node which is an common parent for vertices \(c\)
and \(d\) the second shows conditioning on a node which is a common
child of vertices \(c\) and \(d\). We note that following
\cite{pearl1995} since vertices \(c\) and \(d\) are colliders,
conditioning on them leads to dependence so that
\(a \not\perp d \vert \lbrace b,c \rbrace\) and
\(b \not\perp a \vert \lbrace a,d \rbrace\) in Figure \ref{fig:DAG_1a}.
Figure \ref{fig:DAG_1b} show a DAG where we have \(a\not\perp b\) as
there is conditioning on a common child which is a collider.

\label{fig:DAG_1}

\section{Four Node Chain Graph Example}

\begin{figure}
\begin{center}
\includegraphics[width=\textwidth]{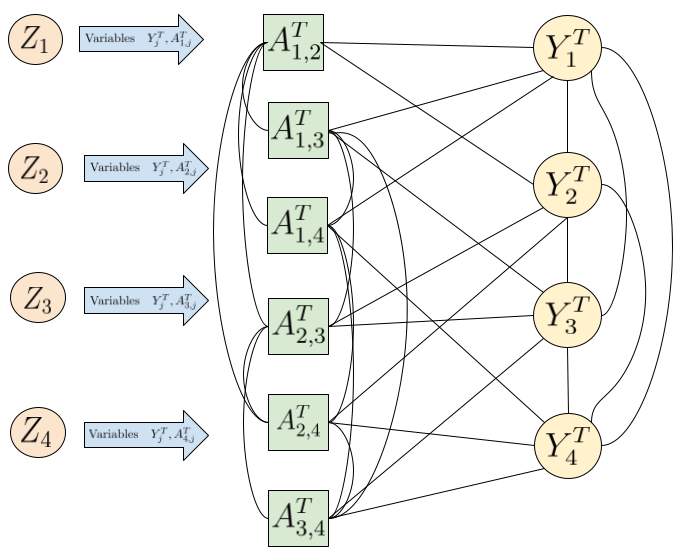}
\caption{Chain graph approximation of a 4 node temporal social network process with pre process treatment and the Markov assumption for edge dependence}
\label{fig:4_node_chain}
\end{center}
\end{figure}

Figure \ref{fig:4_node_chain} shows a chain graph approximation to the
true DAG for a four node social process. Although there are six edge
variables \(A_{i,j}\), each edge variable is only connected to four
other edges. In general the full deletion of connections in the chain
graph is regarded as a very restrictive assumption on the network
dependence structure in the social networks literature. We note that for
networks of larger size the proportion of connections missing compared
to the complete chain graph grows. Thus for networks of size in the
hundreds of nodes, as in our examples, the chain graph approximation
strongly restricts the full dependence structure. It is this restriction
that allows for the modelling of the complete network.

\section{Exponential-family Random Network Models}\label{sec:modeling}

In this section we briefly introduce the Exponential-family Random
Network model \citep{fellows2012} and explain the exchange algorithm
\citep{Murray2006}. This section gives a brief introduction to a complex
method, this paper uses the ERNM and exchange algorithm machinery,
though it is not the primary purpose, so the unfamiliar reader should
consult the above citations. The exchange algorithm is used for sampling
from ERNM posterior distributions, sampling from ERNMs is completely
analogous to the exchange algorithm for ERGM
\citep{Ranran2011, caimo2011}.

The ERNM model can be viewed as a generalization of ERGM to allow for
random nodal covariates, alternatively and equivalently it can be viewed
as generalization of Markov Random Fields (MRF) that allows for random
edges. The basic formulation is an exponential family as follows, for
network \(y\) with nodal covariates \(X\).

\begin{align}
P(Y=y,X=x \vert \eta) = \frac{1}{c(\eta, \mathscr{N})} \exp\left(\eta \cdot g(y,x) \right)
\end{align}

The sample space \(\mathscr{N}\) is the space of all possible binary
edge realizations together with all of the possible random nodes, i.e
\(\mathscr{N} \subset 2^{\mathbb{Y}} \times \mathscr{X}^{n}\), where
\(2^{\mathbb{Y}}\) is the power set of all the dyads and
\(\mathscr{X}^n\) is the joint sample space of the \(n\) nodal
covariates. For formal details see \cite{fellows2012}.

In the ERGM framework it is usual to include statistics that are counts
of specific sub graph realizations, for example the number of triangles
and stars in a network. This accounts for the social structure of many
real networks, which often exhibit more social closure or transitivity
than can be explained by covariates alone. In addition one can include
terms typically included in a MRF models e.g.~the number of neighbours a
node has with a given covariate.

Similar to ERGMs, ERNMs have problems with degeneracy
\citep{handcock2003}. \cite{fellows2012} comment on this, in particular
they propose the use of a somewhat unusual term for homophily which is
usually required to be able to fit ERNMs. We use this term in our
examples, as well as geometrically weighted edgewise shared partner and
degree terms \citep{snijders2006} to allow us to fit models that are not
degenerate.

ERGMs have well defined dependence structures due to the
Hammersley-Clifford Theorem \citep{besag1974}. The dependence structure
of an ERGM is directly related to the choices of sufficient statistics.
A version of the Hammersley-Clifford Theorem is applicable to ERNM
models, though we do not prove it here. Thus, though we consider a close
to complete chain graph to represent the true dependence structure of
the network, we implicitly impose strict dependence structure on the
edges of the network, through our choice of statistics in the ERNM.
However in practice the required choice of homophily statistic for ERNM
renders the dependence structure complex. See \cite{snijders2006} for a
discussion of the dependence structures associated with various
sufficient statistics in the related ERGMs.

Sampling from ERNMs directly is intractable due to the normalizing
constant \(c(\eta,\mathscr{N})\) which is in fact a astronomically high
dimensional sum over the space of all possible graph realizations
\(\sum_{(y,x) \in 2^{\mathbb{Y}} \times \mathscr{X}^{n} } \exp(\eta \cdot g(y,x))\).
As with ERGMs an MCMC routine is required to sample from the
distribution.

ERNMs also have the so called ``doubly intractable'' property that both
the likelihood and the posterior contain intractable normalizing
constants. This renders a fully Bayesian approach challenging. As a
result we use MCMC sampling with the exchange algorithm analogously to
the procedure developed for ERGM. That is instead of sampling from the
distribution \(p(y,x \vert \eta)\), we sample from an augmented
distribution as following. Going forward we abuse notation and denote
\((y,x)\) and simply \(y\), here \(y\) are now realizations of random
edges and nodal covariates.

\begin{align}
p(\eta^{\prime},y^{\prime},\eta\vert y_{obs}) \propto p(y_{obs} \vert \eta)p(\eta^{\prime}\vert \eta)p(y^{\prime} \vert \nu^{\prime})
\end{align}

The idea is to at each step propose a \(\nu^{\prime}\) from some
proposal distribution \(q\), simulate \(y^{\prime}\) from
\(p(y^{\prime}\vert \eta^{\prime})\) and then accept \(\eta^{\prime}\)
with probability:

\begin{align}
\alpha = min(1,\frac{p(\eta^{\prime})p(y_{obs}\vert \eta^{\prime}) q(\eta\vert\eta^{\prime})p(y^{\prime}\vert \eta)}{p(\eta)p(y_{obs}\vert \eta) q(\eta^{\prime}\vert\eta)p(y^{\prime}\vert \eta^{\prime})})
\end{align}

We note that simulating \(y^{\prime}\) from
\(p(y^{\prime}\vert \eta^{\prime})\) is not simple, as sampling from an
ERNM itself requires an MCMC procedure, thus the exchange algorithm for
ERNM is computationally demanding.

We note that often the challenge with achieving plausible convergence to
the stationary distribution with this method has been framed as a
difficulty in choosing a ``good'' the proposal distribution
\(q(\eta^{\prime} \vert \eta)\) \citep{caimo2011}. Here ``good'' in this
case relates to the number of steps before the claim of reaching the
stationary distribution is made, though of course in infinite time
generic proposal step will theoretically lead to convergence. Indeed
naive approaches to adapting the exchange algorithm to ERGMs result in
inability to achieve so called ``burn-in''. Poor proposals can be viewed
as proposals that produce extreme values of
\(\frac{p(y^{\prime}\vert \eta)}{p(y^\prime \vert \eta^{\prime})}\)
either extremely large, or close to \(0\).

We believe this is often observed as a consequence of the non obvious
geometry of the parameter space, coupled with the degenerate nature of
ERNMs outside areas of high probability. Outside plausible areas the
models produce pathological graphs, for example full or empty graphs,
and therefore such proposals are rarely accepted. However if the sampler
does ever end up or even start in these locations, it often rarely
escapes in a practical time. Intuitively we imagine this as the shape of
the posterior parameter space being like a knife edge ridge, to explore
we need to move back and forth along the ridge. The procedure will
rarely accept a proposal falling far of the ridge, but if we get close
to the edge, the sampler can easily fall off and then rarely be able to
get back on to the ridge. Sampling naively, is like trying to explore
the ridge without any knowledge of how wide it is, inevitably the
intrepid explorer falls off and is lost for ever in low posterior
probability land.

The theoretically natural approach to accounting for this geometry is to
enforce a strict prior to this effect. The problem is specifying such a
prior is itself intractable. Some efforts have been made to specify
better priors for this problem \citep{Ranran2011}, using conjugate
priors, or so called non-degeneracy priors. However these approaches by
themselves have not yet yielded a practical sampling methodology.

With the absence of the ability to specify better priors, there have
been efforts to choose the proposal distribution, so that practical burn
in can be achieved. The current state of the art as proposed in
\cite{Ranran2011} and \cite{caimo2011}, use adaptive techniques to learn
the geometry. However we have not achieved efficient sampling with these
methods in networks in the with one-hundred or more nodes with complex
models and 10 or more parameters.

We found it to be very helpful to let \(q(\eta^{\prime}\vert \eta)\) to
be the multivariate Gaussian centred at \(\eta\) with covariance matrix
\(\Sigma\) set to be a scaled inverse Fisher information matrix with
tuning parameter \(\alpha\) as follows:

\begin{align}
\Sigma &= \alpha\cdot I(\eta)^{-1}\\
       &= \alpha\cdot \left(-\mathbb{E}\left[\left(\frac{\partial}{\partial\eta} log(p(y\vert \eta))\right) \vert \eta   \right]\right)^{-1} \\
       &= \alpha \cdot \left(-\mathbb{E}\left[ \left(g(y) - \mathbb{E}\left[g(y) \right] \right)^{\top} \left(g(y) - \mathbb{E}\left[g(y) \right] \right)\right]\right)^{-1}\\
\end{align}

We can estimate this as the sample version, utilizing the simulations
\(\lbrace y_i \rbrace_{i=1}^{m}\) generated from the MCMC, which we
already had to run to the stationary distribution to generate
\(y^{\prime}\), so there is little additional computational cost.

We do not explore principled methods of tuning \(\alpha\)
\citep{hummel2012}. Typically we found an \(\alpha\) between \(0.1\) and
\(0.5\) to facilitate fast exploration of the space. We found that this
proposal distribution allowed us to sample from the posterior ERNM and
ERGM distributions for our \(869\) node example in the order of minutes.

\section{Goodness of Fit for National Longitudinal Study of Adolescent Health High School Smokers}

In this section we present a goodness of fit analysis for the real data
example of smoking behaviour in adolescents in a high school.

In social network analyses usually only one network is observed, we use
Bayesian posterior prediction goodness of fit graphics in the style of
\cite{Hunter_Goodreau_2008}, where simulated distributions of network
summary statistics are compared to the observed values. The simulations
are derived from the posterior, that is new networks are simulated by
first drawing \(\eta\) from the posterior distribution sample and then
simulating a network with that \(\eta\). The choice of the statistics of
interest is subjective, in the social networks literature degree,
edgewise shared partner, and geodesic distance distributions are often
considered.

\begin{figure}[H]

{\centering \includegraphics[width=1\linewidth]{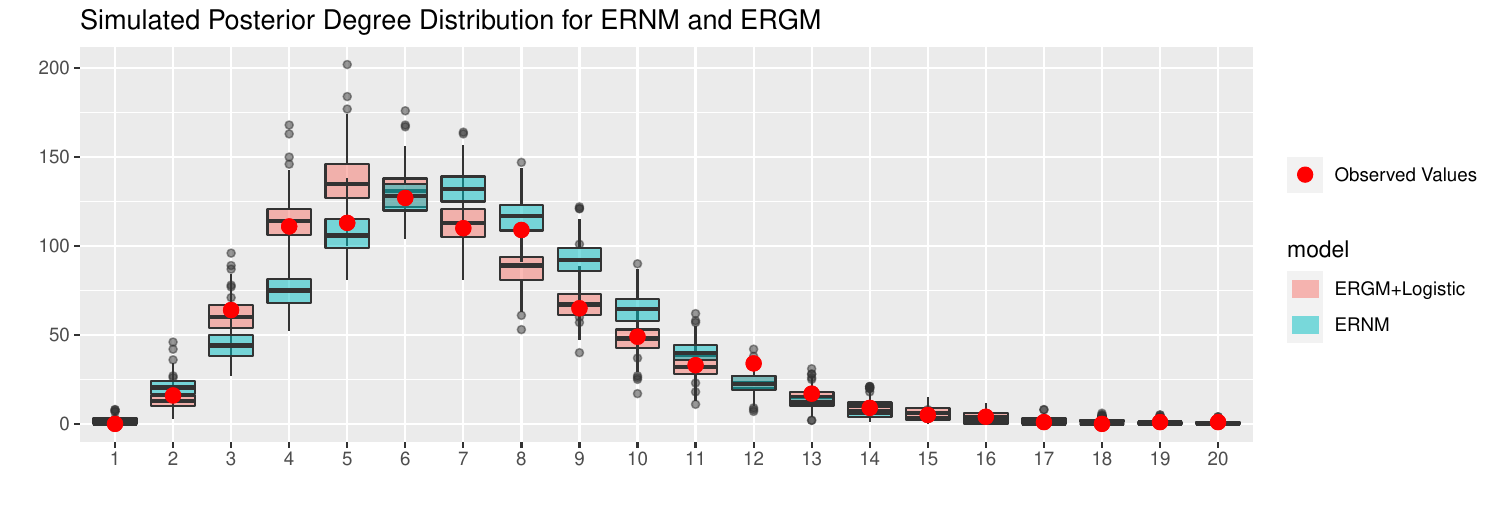} 

}

\caption{\label{fig:gof_degree} ERNM and ERGM simulated posterior degree distribution}\label{fig:unnamed-chunk-2}
\end{figure}

\begin{figure}[H]

{\centering \includegraphics[width=1\linewidth]{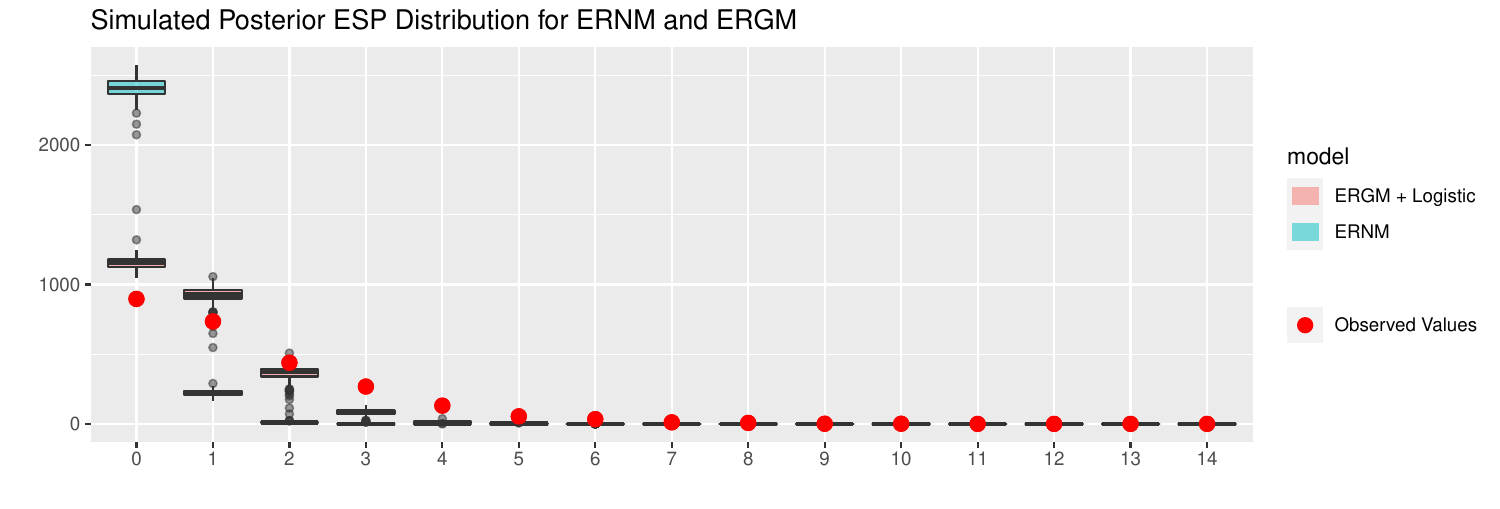} 

}

\caption{\label{fig:gof_esp} ERNM and ERGM simulated posterior ESP distribution}\label{fig:unnamed-chunk-3}
\end{figure}

\begin{figure}[H]

{\centering \includegraphics[width=1\linewidth]{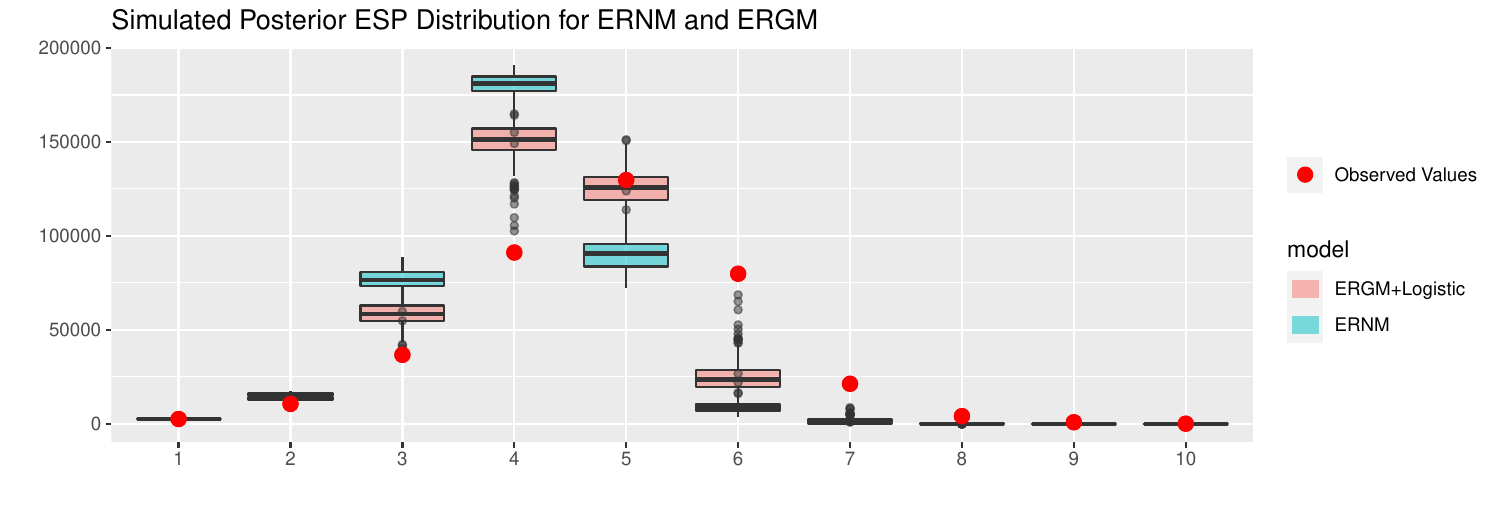} 

}

\caption{\label{fig:gof_geodist} ERNM and ERGM simulated posterior geodesic distance distribution}\label{fig:unnamed-chunk-4}
\end{figure}

Figures \ref{fig:gof_degree}, \ref{fig:gof_esp} and
\ref{fig:gof_geodist} compare the fits of the ERNM and ERGM models, for
the degree, ESP and geodesic distance distributions respectively. In the
case of the logistic regression and MRF models, we do not assess the
goodness of fit of summary statistics involving edges only, as the edges
are not considered random. We note that both models do not fit very well
on ESP, but do well on the degree distribution. In our experience the
fit on geodesic distance distribution is comparable to fits often
observed when fitting these models to social network data.

We note that all these models are in fact highly parsimonious for a
complex social process and suffer from degeneracy issues when additional
terms are added, thus we expect achieving good fit to be very
challenging. It may be the case that tapering \citep{Fellows2017}, which
would allow for more realistic terms without degeneracy may be required
to obtain a well fitting model.

\begin{figure}[H]

{\centering \includegraphics[width=1\linewidth]{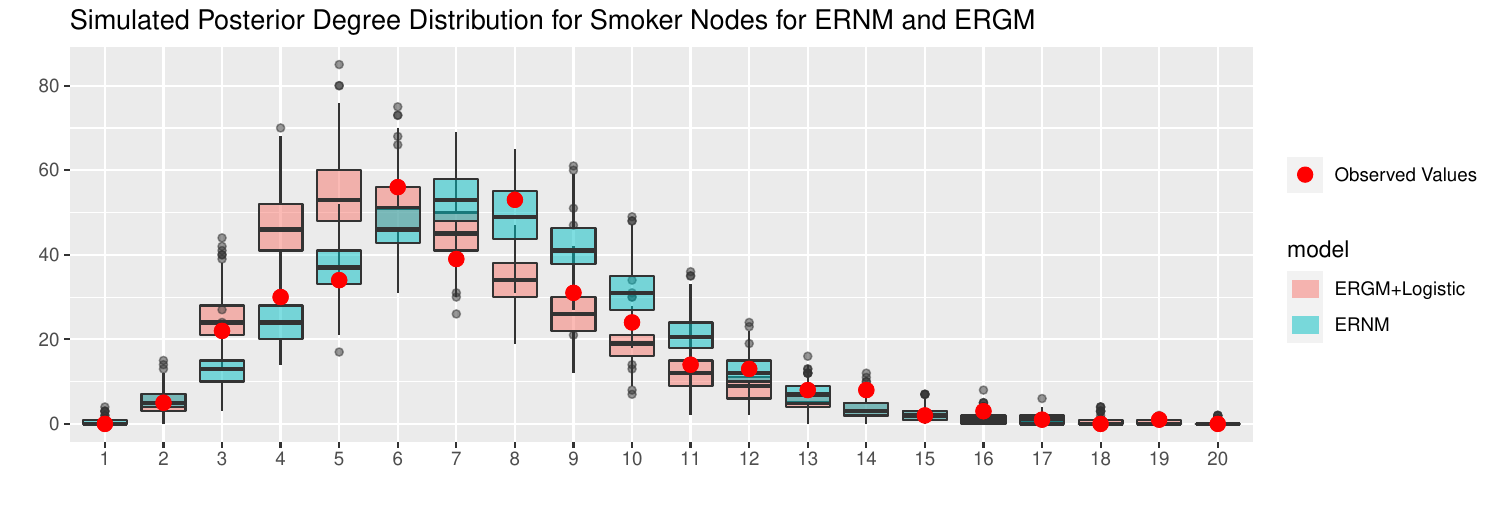} 

}

\caption{\label{fig:gof_degree_smoker} ERNM and ERGM simulated posterior degree distribution for smoker nodes only}\label{fig:unnamed-chunk-5}
\end{figure}

\section{MCMC Convergence Checks for National Longitudinal Study of Adolescent Health High School Smokers Network}

\begin{figure}[H]

{\centering \includegraphics[width=\textwidth]{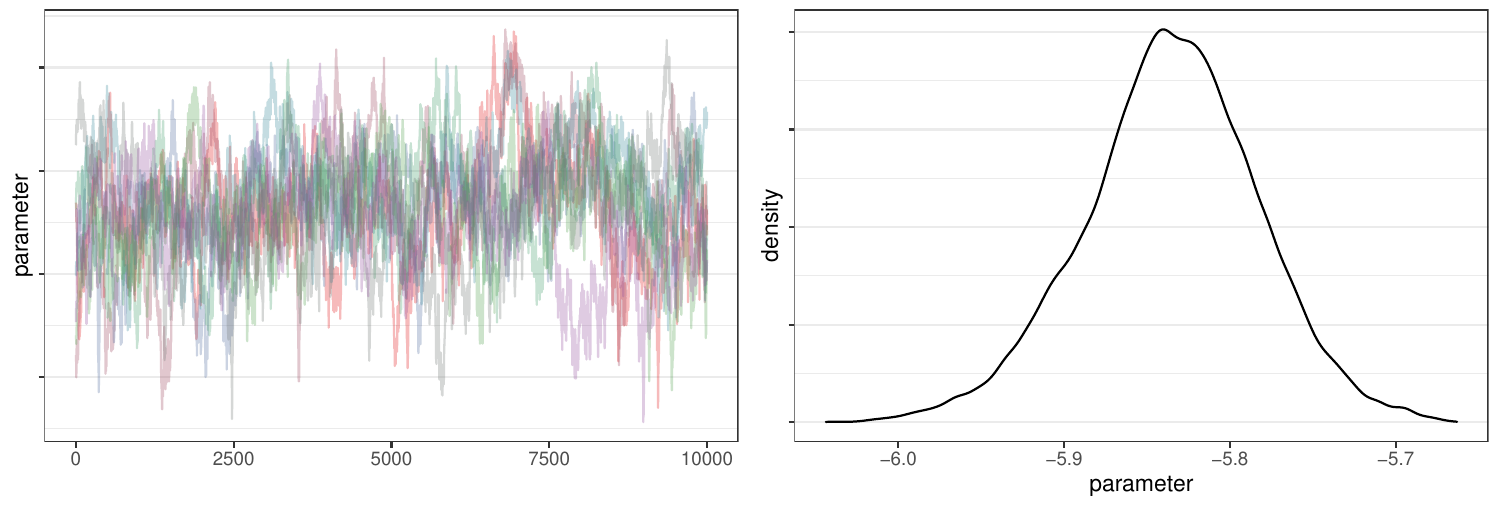} 

}

\caption{\label{fig:trace_edge} Trace plot for edge parameter}\label{fig:unnamed-chunk-7}
\end{figure}

\begin{figure}[H]

{\centering \includegraphics[width=\textwidth]{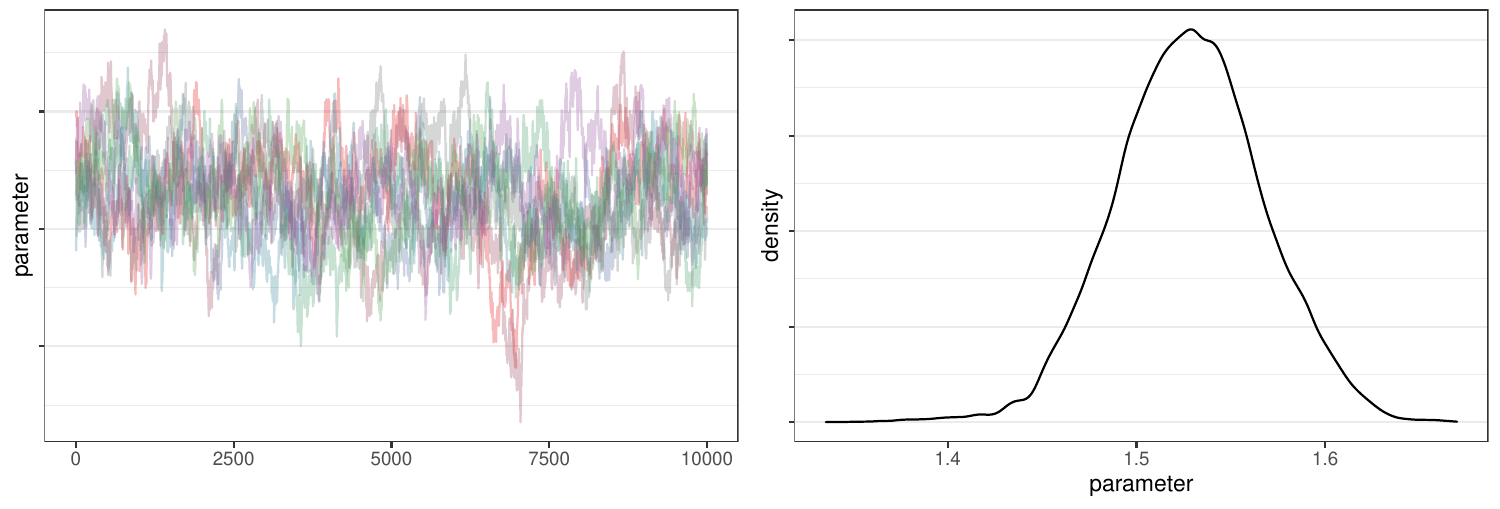} 

}

\caption{\label{fig:trace_GWESP} Trace plot for GWESP parameter}\label{fig:unnamed-chunk-8}
\end{figure}

\begin{figure}[H]

{\centering \includegraphics[width=\textwidth]{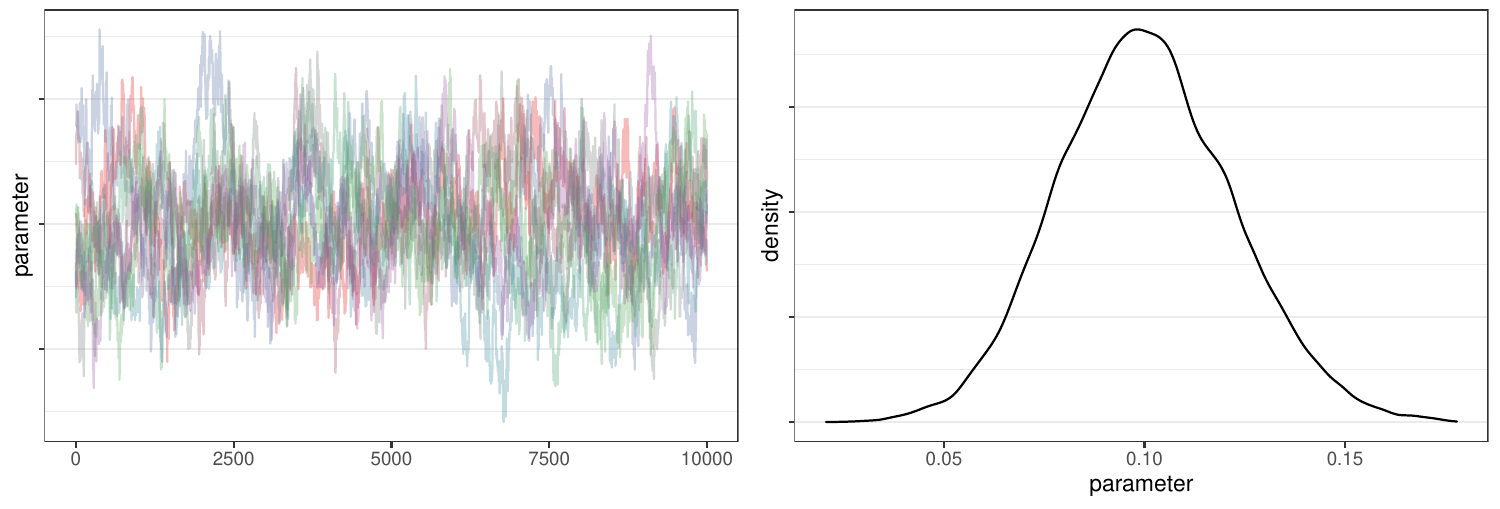} 

}

\caption{\label{fig:trace_smoker_neighbor} Trace plot for smoker neighbor parameter}\label{fig:unnamed-chunk-9}
\end{figure}

We examine the trace plots of the MCMC procedure. Figures
\ref{fig:trace_edge}, \ref{fig:trace_GWESP} and
\ref{fig:trace_smoker_neighbor} show the trace plots for the edge, GWESP
and smoker neighbours terms respectively. We can see that after around
2500 proposals for these parameters the chains appear to vary
independently of their starting point, suggesting the convergence of the
MCMC procedure. We show only three parameters for brevity.

\begin{figure}[H]

{\centering \includegraphics[width=\textwidth]{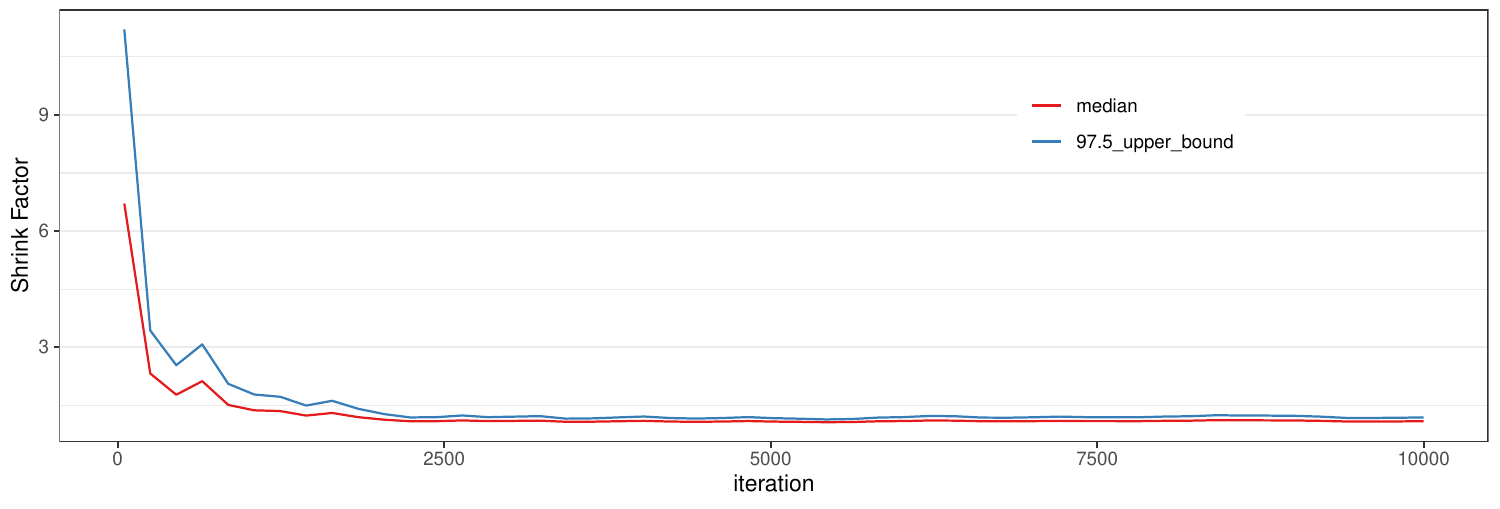} 

}

\caption{\label{fig:gelman_edge}Plot of Gelman diagnostics for edge parameter}\label{fig:unnamed-chunk-11}
\end{figure}

\begin{figure}[H]

{\centering \includegraphics[width=\textwidth]{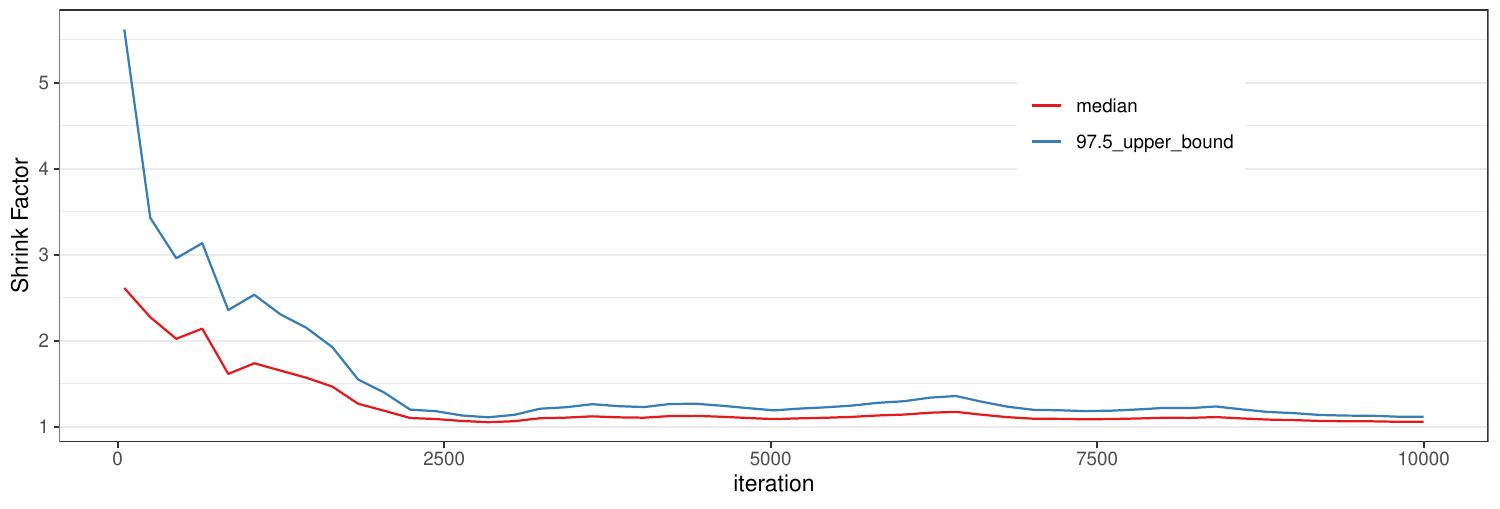} 

}

\caption{\label{fig:gelman_gwesp}Plot of Gelman diagnostics for GWESP parameter}\label{fig:unnamed-chunk-13}
\end{figure}

\begin{figure}[H]

{\centering \includegraphics[width=\textwidth]{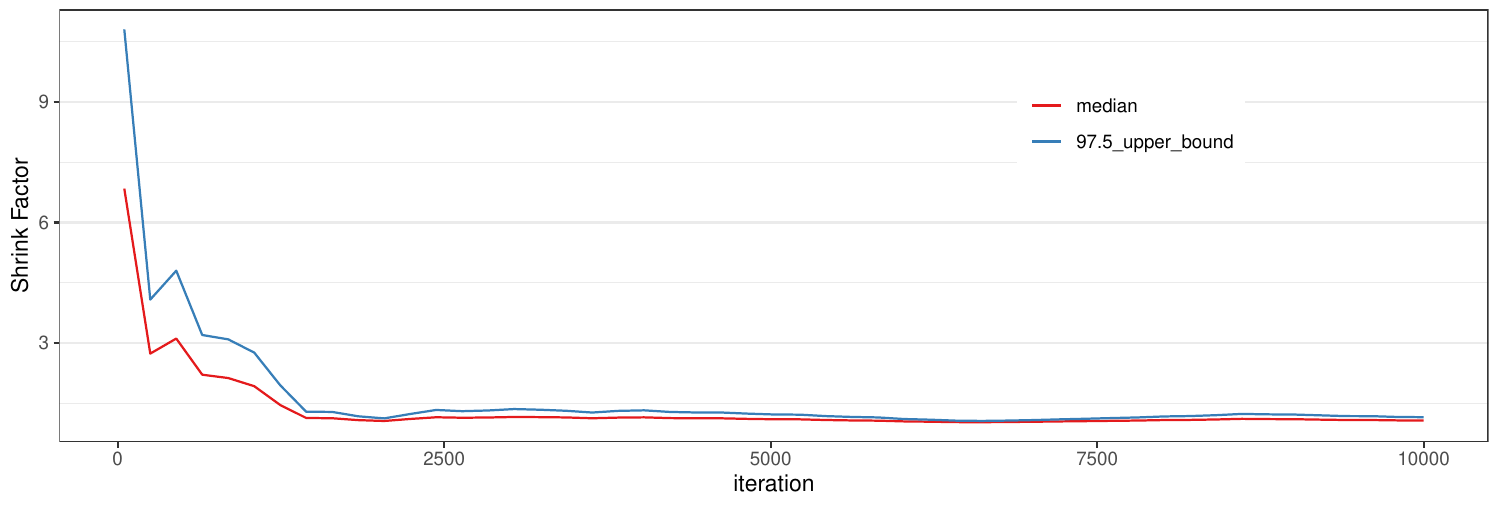} 

}

\caption{\label{fig:gelman_smoker_neighbor} Plot of Gelman diagnostics for smoker neighbors parameter}\label{fig:unnamed-chunk-15}
\end{figure}

We also follow the approach in \cite{Gelman1992} and consider the ratio
of the between chain variance to the within chain variance for \(8\)
MCMCs. A ratio of close to \(1\) suggests that the chain has converged
to its stationary distribution. Figures \ref{fig:gelman_edge},
\ref{fig:gelman_gwesp} and \ref{fig:gelman_smoker_neighbor} show the
ratio plots for the edge, GWESP and smoker neighbours terms
respectively.

\section{Gender Peer Effects for Addolescent Health Network}

We consider the k-peer-gender effect on the ego smoking of having \(k\)
additional friends has over having no friends of specified gender. We
note that the causal effect of gender is controversial to interpret,
however considering say the \(k\)-female-peer effect, may prove
insightful, as one of the consequences of being female is a
significantly higher liklihood of having many female friends.

\begin{landscape}
\begin{figure}[H]

{\centering \includegraphics{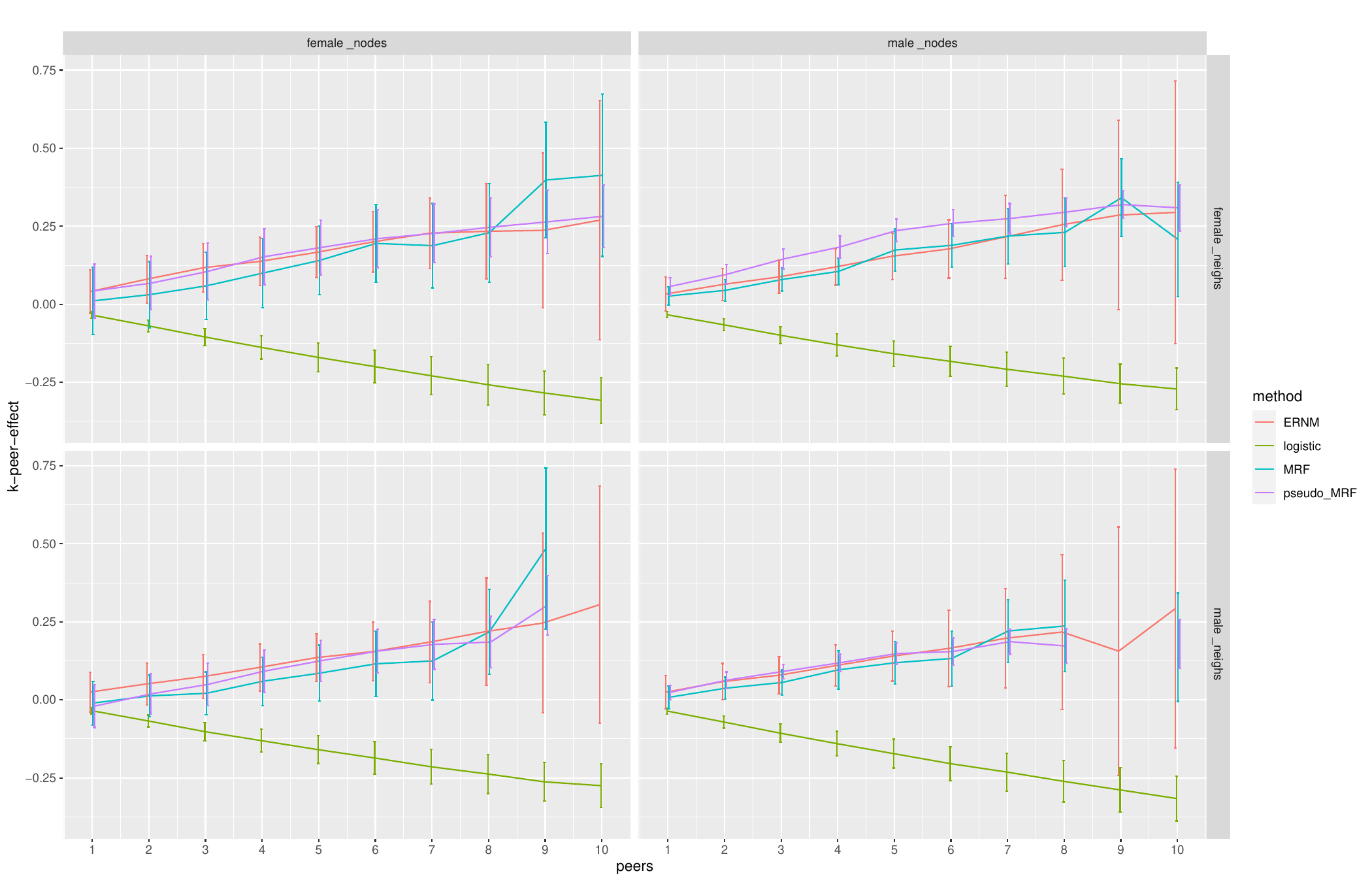} 

}

\caption{\label{fig:big_gender_plot} Plot of k-gender effects for Addolescent Health network, split by node gender and neighbour gender}\label{fig:unnamed-chunk-16}
\end{figure}
\end{landscape}

Figure \ref{fig:big_gender_plot} shows the estimated effects, where the
gender of the nodes and gender of the neighbours considered is varied.
We concentrate on the female neighbours for female nodes, and male
neighbours for male nodes, since there is strong homophily on gender,
these outcomes are most likely to occur in the social process that
generated the network. Interventions that set a node to have more
friends of the opposite gender are likely to need to be extreme to
counteract the overall social forces, inherent in the network formation
process.

\begin{landscape}
\begin{figure}[H]

{\centering \includegraphics{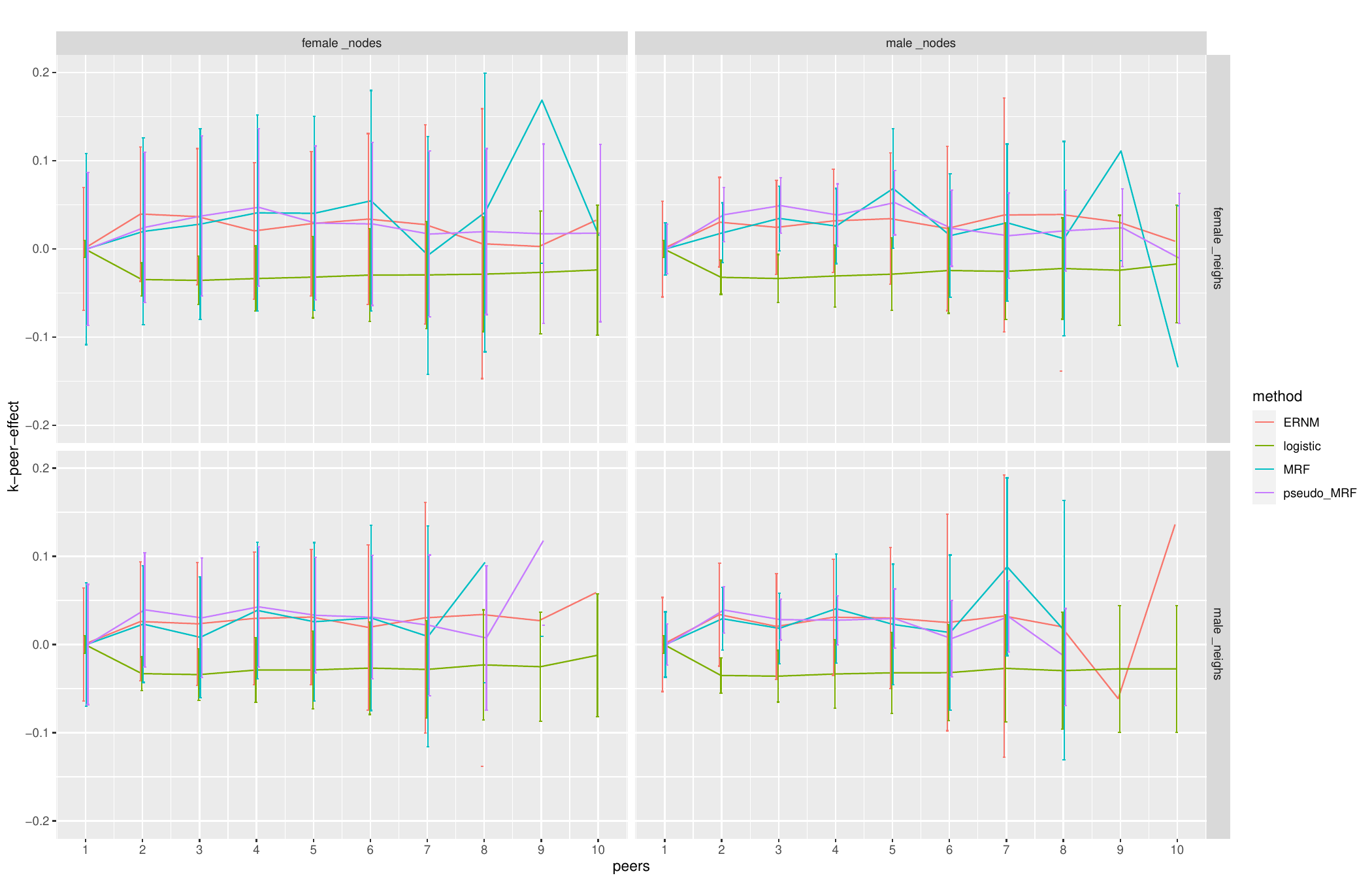} 

}

\caption{\label{fig:big_gender_marginal_plot} Plot of k-gender marginal effects for Addolescent Health network, split by node gender and neighbour gender}\label{fig:unnamed-chunk-17}
\end{figure}
\end{landscape}

We note that little difference is shown between the effect of male
neighbours on male nodes and the effect of female neighbours on female
nodes. We suggest that this shows the effect of more neighbours in
general is the driving force behind increased likelihood of smoking
adoption among nodes with many friendships. Figure
\ref{fig:big_gender_marginal_plot} shows the is little difference in the
marginal increase in effect from each additional friend under all
models.

Clearly the logistic regression model performs poorly here, suggesting
that more neighbours has a negative effect on smoking. This is due to
the confounding of the number of female neighbours, with the number of
smoker neighbours, which results in a negative estimate for this
coefficient. We suggest that this could be mitigated with a different
specification, allowing for interactions, however we include this model
as precautionary tale against using naive models for network data. The
ERNM and MRF are statistically indistinguishable, whilst the pseudo MRF
also seems similar.

\begin{landscape}
\begin{figure}[H]

{\centering \includegraphics{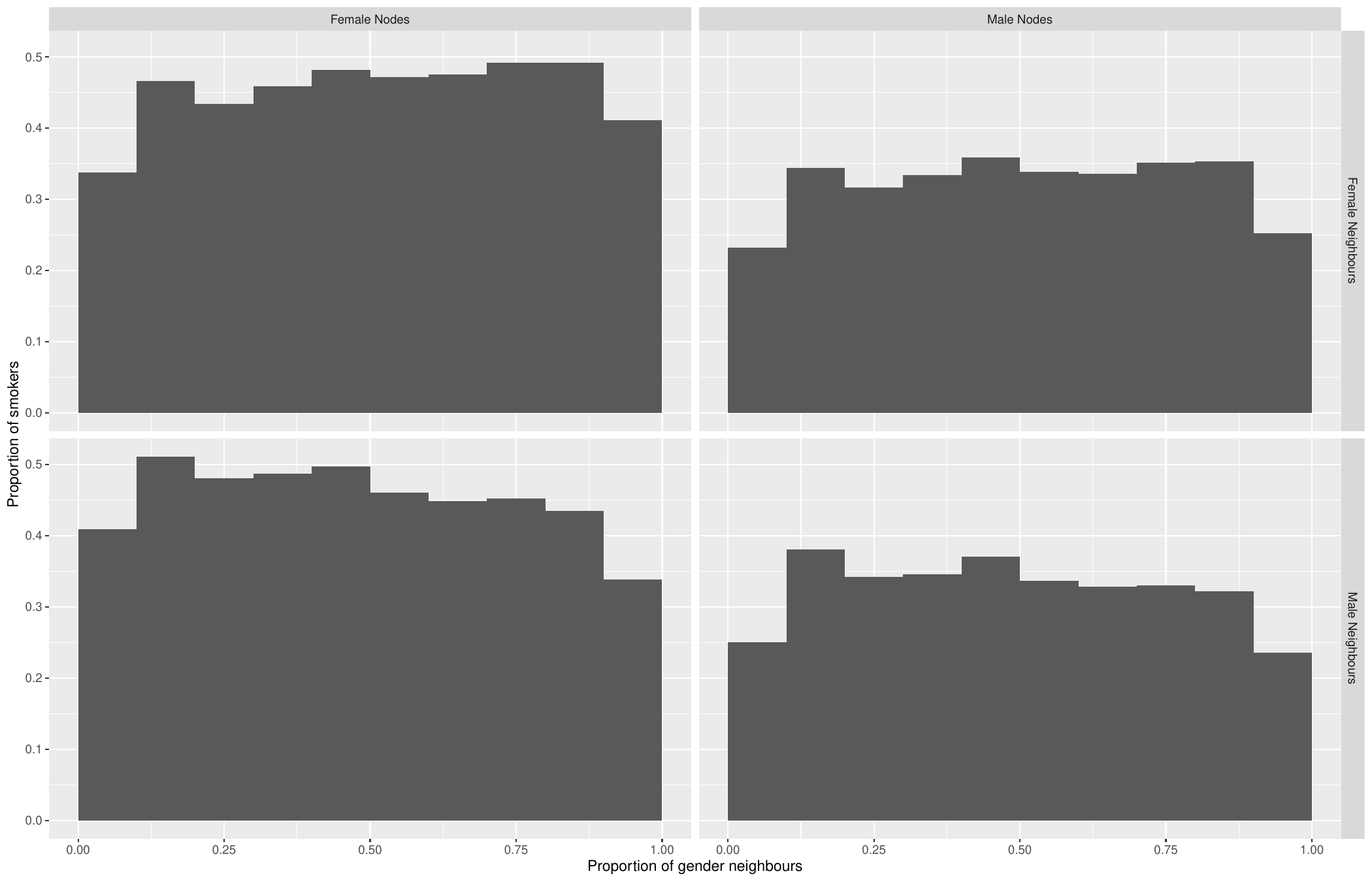} 

}

\caption{\label{fig:prop_homo_friends} ERNM model barplot of mean smoker status, for each proportion of neighsbours of each gender, for each ego gender. We see that the proportion of neighbors of either particular gender does notimpact the likelihood of smoking.}\label{fig:unnamed-chunk-18}
\end{figure}
\end{landscape}

However as the raw number of friendships a female node has is highly
correlated with the number of female friendships a node has, we caution
interpreting the effect, as that of gendered friends, rather that of
friends in general. In particular Figure \ref{fig:prop_homo_friends}
shows the effect of the proportion of genders an egos friends are for
the ERNM model. The proportion of an egos friends that are homogeneous
on gender does not seem to greatly effect the likelihood of smoking. We
note the top right panes on this have lower overall smoking levels as
observed for male nodes in the data. This shows that there is limited
evidence that the gender of a nodes friends has an effect on the
likelihood of smoking.

\newpage

\newpage

\bibliographystyle{chicago}
\bibliography{bib_causal}